\documentclass[letterpaper,journal]{IEEEtran}
\usepackage{algorithm}
\usepackage{algorithmic}
\usepackage{mathrsfs}
\usepackage[T1]{fontenc}
\usepackage{textcomp}
\usepackage{cite}
\usepackage{latexsym}
\usepackage{amsmath,amssymb,amsfonts}
\usepackage{newtxtext}
\usepackage{newtxmath}
\usepackage{bm}
\usepackage{url}
\usepackage{mathtools}
\usepackage{comment}
\usepackage{graphicx}
\usepackage{subfigure}
\usepackage{xcolor}
\usepackage{overpic}
\usepackage{epstopdf}
\usepackage{float}
\usepackage{pdfpages}
\usepackage{svg}
\usepackage{array}
\newcommand{\red}[1]{{\color{black}{#1}}}
\newcommand{\redC}[1]{{\color{black}{#1}}}
\newcommand{\tealA}[1]{{\color{black}{#1}}}

\begin{document}

\title{Joint Optimization of Antenna Partitioning and Distortion-Aware Precoding for Massive MIMO-OFDM with Peak Cancellation and In-Band Distortion Compensation}

\author{Zhuoran~Li,~\IEEEmembership{Graduate~Student~Member,~IEEE},
        Osamu~Muta,~\IEEEmembership{Member,~IEEE},
        and~Haris~Gacanin,~\IEEEmembership{Fellow,~IEEE}
\thanks{This work has been submitted to the IEEE for possible publication. Copyright may be transferred without notice, after which this version may no longer be accessible.}%
\thanks{(Corresponding author: Zhuoran Li.)}%
\thanks{Zhuoran Li is with the Graduate School of Information Science and Electrical Engineering, Kyushu University, Fukuoka 819-0395, Japan (e-mail: z.li@mobcom.ait.kyushu-u.ac.jp).}%
\thanks{Osamu Muta is with the Faculty of Information Science and Electrical Engineering, Kyushu University, Fukuoka 819-0395, Japan (e-mail: muta@ait.kyushu-u.ac.jp; muta@ieee.org).}%
\thanks{Haris Gacanin is with the Chair for Distributed Signal Processing, RWTH Aachen University, 52074 Aachen, Germany (e-mail: harisg@dsp.rwth-aachen.de).}%
}

\markboth{Journal of \LaTeX\ Class Files,~Vol.~14, No.~8, August~2021}%
{Shell \MakeLowercase{\textit{et al.}}: A Sample Article Using IEEEtran.cls for IEEE Journals}


\maketitle

\begin{abstract}
In massive MIMO-OFDM systems, peak cancellation (PC) reduces the
peak-to-average power ratio but introduces in-band distortion, where 
compensation signals transmitted from a subset of antennas can suppress this distortion at the users. 
However, their effectiveness depends on the antenna partition and compensation precoder, 
while nonlinear power amplifiers (PAs) generate additional distortion. 
This paper proposes a joint optimization method for antenna partitioning and distortion-aware compensation precoding. 
To enable this joint optimization, we derive 
a covariance-based analytical model of PC and PA distortion to predict the per-user, per-subcarrier signal-to-interference-plus-distortion-and-noise ratio (SIDNR). 
The predicted SIDNR is used to select the compensation precoder regularization, the disjoint antenna partition, and the number of compensation antennas. 
Numerical results demonstrate that joint optimization of the antenna partition and compensation precoder regularization improves BER and SIDNR-based spectral efficiency compared with fixed-partition designs, particularly at high signal-to-noise ratios.
\end{abstract}

\begin{IEEEkeywords}
Antenna partitioning, distortion covariance, in-band distortion compensation,
massive MIMO-OFDM, nonlinear power amplifiers, peak cancellation, SIDNR.
\end{IEEEkeywords}

\section{Introduction}
Massive multiple-input multiple-output (mMIMO) combined with orthogonal
frequency-division multiplexing (OFDM) supports multiuser broadband
transmission, but its high peak-to-average power ratio (PAPR) complicates
power-amplifier (PA) operation~\cite{RahmatallahMohan2013}.
Operating a PA near saturation introduces nonlinear distortion, which can cause both in-band distortion and out-of-band radiation, whereas
increasing the input back-off reduces power efficiency~\cite{Ochiai2013}.
This tradeoff is particularly important at a mMIMO base station (BS) with multiple transmit chains~\cite{Zayani2019PAPR}.

\redC{Peak cancellation (PC) suppresses waveform peaks by adding cancellation
pulses, without side information or additional receiver processing, but
stronger peak suppression introduces greater in-band
distortion~\cite{SongOchiai2016,tvtKageyama}. In mMIMO, this distortion
can be mitigated at the users by exploiting the spatial redundancy of the array~\cite{Zayani2019PAPR}. To this end, the antennas are partitioned into disjoint
data and compensation subsets, and the latter transmits precoded
compensation signals, which allows more aggressive peak
suppression~\cite{KageyamaAccess2020}. However, the compensation waveform is not peak-limited and may drive the compensation PAs into nonlinear operation, so the compensation precoder must balance the residual PC distortion against the PA distortion caused by the compensation signal itself. This tradeoff motivated distortion-aware minimum mean-square error (MMSE) compensation precoding~\cite{Li_VTC}. Moreover, the partition
determines the effective channels of both branches and the number of data
antennas, and the compensation-antenna placement affects the
compensation-signal peaks and the received
performance~\cite{Li_ACCESS,Li_IEICE}. The partition and the precoder are thus coupled: a poorly conditioned compensation channel, e.g., from closely spaced compensation antennas, produces a strong and peaky compensation signal unless the precoder is regularized accordingly. These studies, however, treated antenna
placement and compensation precoding separately. 
A joint design requires a tractable model that relates the partition and the precoder to
the received performance under PC and PA distortion. Such a model is needed because
waveform-level simulation of every candidate is computationally demanding, particularly
when the number of compensation antennas is also variable.}

This paper proposes a joint optimization method for antenna partitioning
and distortion-aware compensation precoding in massive MIMO-OFDM systems
with peak cancellation and in-band distortion compensation at a transmitter with nonlinear PAs. 
To enable this joint optimization, we derive a covariance-based analytical signal-to-interference-plus-distortion-and-noise ratio (SIDNR) model 
and use it to evaluate the candidate designs.
The main contributions are as follows.

\redC{
\begin{itemize}
  \item We derive a per-user, per-subcarrier approximation of the SIDNR,
  accounting for multiuser interference, residual PC distortion, PA
  distortion from both transmit branches, and receiver noise.

  \item Based on the analytical SIDNR model, we formulate and solve a constrained optimization problem for maximizing the average spectral efficiency by jointly optimizing the antenna partition, compensation precoder regularization, and number of compensation antennas.
\end{itemize}
}

\noindent\textbf{Notation:}
Vectors and matrices are denoted respectively by bold lowercase and uppercase
letters. The set of complex matrices with dimensions $a\times b$ is denoted by
$\mathbb C^{a\times b}$. The operators $(\cdot)^\mathrm T$ and
$(\cdot)^\mathrm H$ denote transpose and Hermitian transpose, respectively.
The Kronecker product is denoted by $\otimes$. Other notation is summarized in
Table~\ref{table:kigou}.
\begin{table*}[t]
  \centering
  \caption{Principal Notation}
  \label{table:kigou}
  \footnotesize
  \setlength{\tabcolsep}{3pt}
  \renewcommand{\arraystretch}{1.05}
  \!\!\begin{tabular}{
    |>{\raggedright\arraybackslash}p{0.06\textwidth}
    |>{\raggedright\arraybackslash}p{0.40\textwidth}
    |>{\raggedright\arraybackslash}p{0.1\textwidth}
    |>{\raggedright\arraybackslash}p{0.40\textwidth}|}
    \hline
    \textbf{Symbol} & \textbf{Description}
    & \textbf{Symbol} & \textbf{Description} \\
    \hline

    $U,\;M$
    & Numbers of users and BS antennas.
    & $\mathcal M,\;\mathcal M_{\mathrm d},\;
       \mathcal M_{\mathrm e}$
    & Full, data, and compensation antenna-index sets. \\
    \hline

    $M_{\mathrm d},\;M_{\mathrm e}$
    & Numbers of data and compensation antennas.
    & $\mathbf S_{\mathrm d},\;\mathbf S_{\mathrm e}$
    & Data- and compensation-antenna selection matrices. \\
    \hline

    $N,\;\mathcal L$
    & DFT size and occupied-subcarrier index set,
      respectively.
    & $\mathbf W_{\kappa,0},\;\mathbf W_\kappa$
    & Unnormalized and applied branch precoders. \\
    \hline

    $\mathbf H,\;\mathbf H_\kappa$
    & Full-array and branch channel matrices.
    & $\mathbf d,\;\bm\eta_{\mathrm{pc}}$
    & Actual PC signal and
      input-uncorrelated Bussgang residual. \\
    \hline

    $\lambda_\kappa,\;\alpha_\kappa$
    & Branch RZF regularization and
      precoder normalization factor.
    & $\bm\epsilon,\;\bm\epsilon_\kappa$
    & Full-array and branch PA-error vectors. \\
    \hline

    $\mathbf x,\;\mathbf x'$
    & Data-branch signals before and after PC.
    & $\mathbf C_{\mathbf x},\;\mathbf C_{\mathbf d}$,
      $\mathbf C_{\mathbf\epsilon,\kappa}$
    & Covariances of the data signal, PC signal, and branch PA error. \\
    \hline

    $\mathbf z,\;\mathbf v$
    & Compensation signal and full-array digital
      PA-input vector.
    & $\sigma_s^2,\;\sigma_n^2$
    & Per-stream symbol power and per-user noise power. \\
    \hline

    $A_{\mathrm{th}},\;A_0$
    & PC threshold and its fixed value
      in the practical design.
    & $\sigma_{x,m}^2,\;\sigma_{v,m}^2$
    & Per-antenna time-domain powers of the pre-PC
      data signal and PA input. \\
    \hline

    $r_{\max},\;\xi_m$
    & PA saturation-reference input amplitude
      and per-antenna linear IBO.
    & $\sigma_{\mathrm{pc},m}^2,\;
       \sigma_{\mathrm{pa},m}^2$
    & Per-antenna time-domain powers of the PC
      cancellation signal and PA error. \\
    \hline

    $P_{\mathrm d},\;P_{\mathrm e}$
    & Total PA-input powers of the data
      and compensation branches.
    & $P_{\max},\;P_{\mathrm{cth}}$
    & Total PA-input power budget and
      compensation-branch power limit. \\
    \hline

    $K,\;\mathcal K$
    & Compensation-antenna count $K=M_{\mathrm e}$
      and its candidate set.
    & $\Gamma_u,\;\widetilde{\Gamma}_u,\;
       \widetilde{\Gamma}_u^{\mathrm r}$
    & Full, practical-design, and relaxed
      SIDNRs of user $u$. \\
    \hline

    $\mathbf q$
    & Relaxed antenna-role vector:
      $q_m=0$ for data and $q_m=1$ for compensation.
    & $R,\;\widetilde R,\;\widetilde R^{\mathrm r}$
    & Full, practical-design, and relaxed
      SIDNR-based rate metrics. \\
    \hline
  \end{tabular}

  \vspace{2pt}
  \begin{minipage}{\textwidth}
    \footnotesize
    \raggedright
    Here, $\kappa\in\{\mathrm d,\mathrm e\}$ denotes
    the data or compensation branch.
    The subcarrier index $[l]$ is omitted from
    frequency-domain quantities for brevity.
  \end{minipage}
\end{table*}

\section{Related Works}
\label{Sect:RelatedWorks}
\tealA{PAPR-aware mMIMO-OFDM downlink transmission jointly performs multiuser precoding and PAPR reduction~\cite{Zayani2019PAPR}.}
Such methods establish the value of spatial redundancy,
but do not address a dedicated PC-distortion compensation branch whose antenna
assignment and PA nonlinearity jointly determine the received distortion.

Analytical PC models commonly exploit the approximate Gaussianity of
time-domain OFDM samples. 
\tealA{Song and Ochiai derived closed-form approximations of the
signal-to-distortion ratio for peak-canceled OFDM signals~\cite{SongOchiai2016}.
In~\cite{tvtKageyama}, the optimum PC threshold was analytically derived
under prescribed error vector magnitude (EVM) and adjacent-channel leakage
ratio (ACLR) constraints.
This threshold determines the achievable peak suppression while keeping
the in-band distortion and out-of-band radiation within these limits.}
However, these analyses do not specify the spatial covariance of PC distortion across transmit antennas or its propagation through compensation precoding. 

\tealA{Enhanced PC with simplified transmitter-side compensation was
proposed in~\cite{KageyamaAccess2020}.
Related antenna-reservation, null-space, and compensation-oriented methods
were investigated
in~\cite{Prabhu2014PAPR,Khan2021TransmitDistortion,Saito2024ParallelPC}.}
Muneer et al.\ examined antenna reservation for clipping-distortion compensation, including the number of reserved antennas and the associated data-array-gain penalty~\cite{Muneer2022AntennaReservation}. Their more recent method instead maps hard-clipping distortion to the user domain, re-precodes it, and superimposes the compensation signal on the clipped data signal at every transmit antenna~\cite{Muneer2026Clipping}. 
\tealA{However, the compensation signal in these methods is itself not peak-limited.
When it is transmitted from the same set of antennas as the peak-reduced signal, the peaks can partly regrow. When it is transmitted from a separate set of reserved antennas, it may drive their PAs into nonlinear operation. In both
cases, the resulting PA distortion depends on which antennas carry the
compensation signal and how it is precoded.}

\tealA{In our previous work, the uniform planar array (UPA)-based scheme in~\cite{Li_ACCESS} showed
that compensation-antenna placement affects spatial correlation,
compensation-signal peak power, and bit error rate (BER).
Placement-aware compensation was extended to subarray-based hybrid mMIMO
in~\cite{Li_IEICE}.}
The distortion-aware MMSE design in~\cite{Li_VTC} further incorporated compensation-branch PA distortion into the
precoder. 
\tealA{However, these studies~\cite{Li_VTC,Li_ACCESS,Li_IEICE} considered
prescribed placements or precoder structures and relied mainly on
waveform-level evaluation.
These works did not provide an analytical SIDNR model that accounts for both PC and PA distortion.
As a result, the partition, the number of compensation antennas, and the precoder were not optimized jointly.}
\redC{Beyond these compensation-oriented studies, antenna configuration has also been investigated in conventional antenna selection, which aims to preserve MIMO gain while reducing radio-frequency (RF) hardware or processing costs by activating only a subset of antennas~\cite{SanayeiNosratinia2004}.}

\tealA{In contrast to the above existing studies, this paper keeps all antennas active and assigns each antenna to either data transmission or PC-distortion compensation.
The partition, the number of compensation antennas, and the
compensation-precoder regularization are jointly selected by using a
covariance-based SIDNR model of PC and PA distortion.}

\section{System Model}
\label{Sect:System Model}

This section presents a description of the considered downlink mMIMO-OFDM system employing PC and in-band distortion compensation at the transmitter. The signal processing steps and system architecture are explained in the order in which they are applied.

\begin{figure*}[tb]
    \centering
    \includegraphics[width=0.9\linewidth]{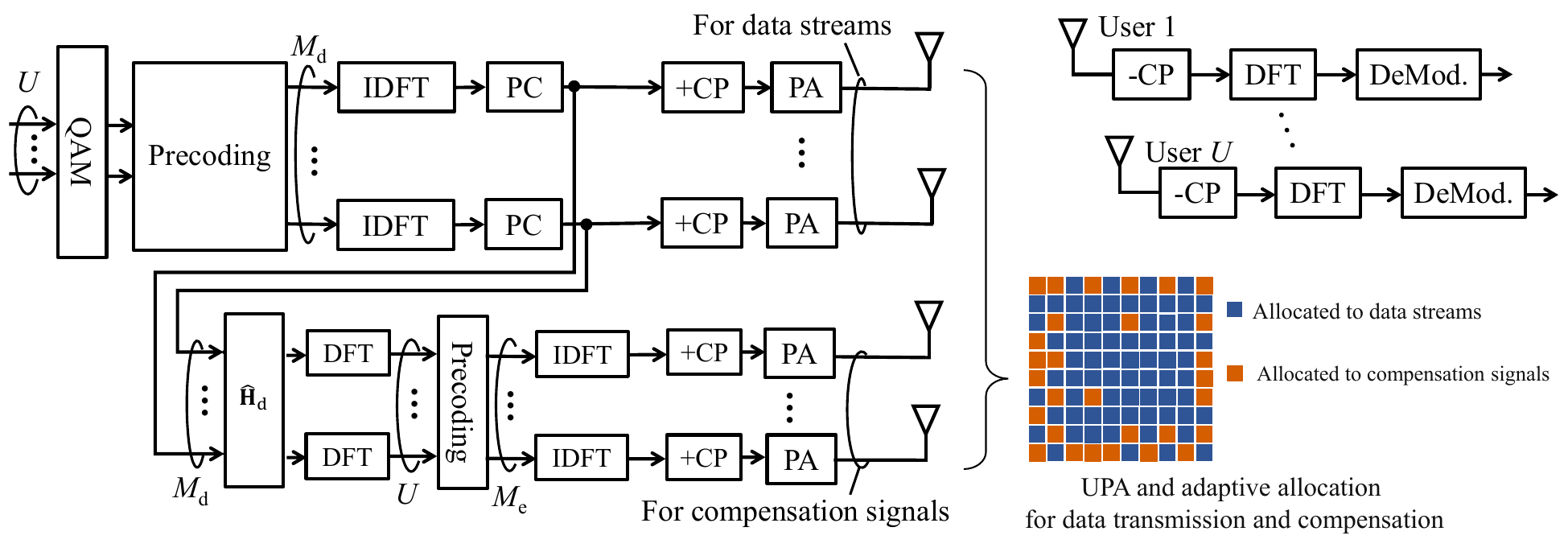}
    \caption{Block diagram of the considered mMIMO-OFDM system with a UPA.}
    \label{fig:system_model}
\end{figure*}

\subsection{System Overview}
As is shown in Fig.~\ref{fig:system_model}, the BS serves $U$ single-antenna users. The BS is equipped with a UPA of $M$ antennas, partitioned into $M_\mathrm{d}$ data antennas for data transmission and $M_\mathrm{e}$ in-band distortion compensation antennas for conveying the in-band distortion compensation signals generated by the PC technique. The relation $M = M_\mathrm{d} + M_\mathrm{e}$ holds. The colored array in Fig.~\ref{fig:system_model}
illustrates one possible partition.

\subsection{Channel Model}
The propagation channel between the BS and the $u$-th user on the $l$-th subcarrier is modeled as~\cite{FDD}:
\begin{equation}
  \mathbf{H}_u[l] = \sum_{p=1}^{N_\mathrm{h}} \zeta_u^p \mathbf{R}_\mathrm{t}^\mathrm{T}(\psi_{\mathrm{t},u}^p) e^{j 2 \pi \tau_u^p f_l},
\end{equation}
where $\zeta_u^p$, $\tau_u^p$, and $\psi_{\mathrm{t},u}^p$ denote the complex path gain, propagation delay, and direction of departure for the $p$-th path of user $u$, respectively; $N_\mathrm{h}$ represents the number of propagation paths; $f_l$ stands for the frequency offset of the $l$-th subcarrier relative to the reference subcarrier, given by $f_l = l \cdot \Delta f$, where $\Delta f$ is the subcarrier spacing. The transmit steering vector $\mathbf{R}_\mathrm{t}(\psi) \in \mathbb{C}^{M \times 1}$ is expressed as the Kronecker product of the horizontal and vertical steering vectors
\begin{equation}
  \mathbf{R}_\mathrm{t}(\psi) = \mathbf{R}_\mathrm{h}(\psi_\mathrm{h}) \otimes \mathbf{R}_\mathrm{v}(\psi_\mathrm{v}),
\end{equation}
where $\psi_\mathrm{h}$ and $\psi_\mathrm{v}$ respectively represent the normalized spatial frequencies corresponding to the physical azimuth angle $\theta$ and elevation angle $\phi$. The horizontal and vertical steering vectors are given as
\begin{align}
\mathbf{R}_\mathrm{h}(\psi_\mathrm{h}) &= \frac{1}{\sqrt{M_\mathrm{ro}}} \left[ 1, e^{-j 2\pi \psi_\mathrm{h}}, \dots, e^{-j 2\pi (M_\mathrm{ro} - 1) \psi_\mathrm{h}} \right]^\mathrm{T}, \nonumber\\
\mathbf{R}_\mathrm{v}(\psi_\mathrm{v}) &= \frac{1}{\sqrt{M_\mathrm{co}}} \left[ 1, e^{-j 2\pi \psi_\mathrm{v}}, \dots, e^{-j 2\pi (M_\mathrm{co} - 1) \psi_\mathrm{v}} \right]^\mathrm{T},
\end{align}
where $M_\mathrm{ro}$ and $M_\mathrm{co}$ respectively denote the numbers of antennas in the horizontal and vertical directions.
Stacking the $U$ user channel vectors, we define the aggregate channel matrix
$
\mathbf{H}[l]
\triangleq
\left[\mathbf{H}^\mathrm T_1[l] , \cdots , \mathbf{H}^\mathrm T_U[l]\right]^{\mathrm T}
\in \mathbb{C}^{U\times M}.
$

\subsection{Digital Precoding}
The digital precoder is designed based on the channel matrix.
Let $\kappa \in \{\mathrm d,\mathrm e\}$ denote the branch index, where
$\kappa=\mathrm d$ corresponds to the data branch and
$\kappa=\mathrm e$ corresponds to the compensation branch.
For subcarrier $l\in\mathcal L$, where $\mathcal L$ is the set of occupied
subcarriers, the effective channel of branch $\kappa$ is denoted by
$\mathbf H_\kappa[l]\in\mathbb C^{U\times M_\kappa}$.
In this paper, we use the scalar regularized zero-forcing (RZF) form~\cite{Peel2005RZF}
\begin{equation}
  \mathbf W_{\kappa,0}[l]
  =
  \mathbf H_\kappa^\mathrm H[l]
  \left(
    \mathbf H_\kappa[l]\mathbf H_\kappa^\mathrm H[l]
    +
    \lambda_{\kappa}\mathbf I_U
  \right)^{-1},
  \qquad \kappa\in\{\mathrm d,\mathrm e\},
  \label{eq:RZF_kappa}
\end{equation}
where $\lambda_{\kappa}\ge 0$ is a scalar regularization parameter.
The applied precoding matrix is then given by
\begin{equation}
  \mathbf W_\kappa[l]=\alpha_\kappa \mathbf W_{\kappa,0}[l],
  \qquad \kappa\in\{\mathrm d,\mathrm e\},
  \label{eq:Wkappa_norm}
\end{equation}
where $\alpha_\kappa$ is the normalization factor for branch $\kappa$.
When $\lambda_{\kappa}=0$, \eqref{eq:RZF_kappa} reduces to the zero-forcing (ZF) precoder.

\subsection{Peak Cancellation}
\label{Sect:Peak Cancellation}
After digital precoding, a PC technique~\cite{tvtKageyama} is applied to reduce the PAPR of the OFDM signal.
PC is a low-complexity method that detects signal samples whose amplitudes exceed a predefined threshold $A_\mathrm{th}$, and then adds specially designed pulses to suppress the peaks down to the threshold level.
\redC{Let $x^\mathrm{td}[n]$ be the time-domain OFDM signal of a given antenna,
and let $n_1,\ldots,n_{N_p}$ be the samples whose magnitudes exceed
$A_\mathrm{th}$. For the peak at $n_i$, the cancellation pulse is
\begin{equation}
  p^\mathrm{td}_{r,i}[n - n_i] = -A_{p,i} e^{j\theta_i} g'[n - n_i],
\end{equation}
where $A_{p,i} = |x^\mathrm{td}[n_i]| - A_\mathrm{th}$,
$\theta_i = \arg\{x^\mathrm{td}[n_i]\}$, and $g'[n]=|\mathcal L|^{-1}\sum_{l\in\mathcal L}e^{j2\pi ln/N}$,
$n=0,\ldots,N-1$, is the full-symbol circular sinc-type kernel
band-limited to the occupied subcarriers, with $g'[0]=1$. The total cancellation signal and the
peak-canceled signal are
\begin{equation}
  d^\mathrm{td}[n] = \sum_{i=1}^{N_p} p^\mathrm{td}_{r,i}[n - n_i],\qquad
  (x')^\mathrm{td}[n] = x^\mathrm{td}[n] + d^\mathrm{td}[n].
  \label{eq:PC4}
\end{equation}
The $N$-point discrete Fourier transform (DFT) gives
\begin{equation}
  x'[l] = x[l] + d[l],
  \label{eq:PC5}
\end{equation}
where $d[l]$, the DFT of $d^\mathrm{td}[n]$, is confined to the occupied
subcarriers. 
Unlike~\cite{tvtKageyama}, which derived the PC threshold from EVM and
ACLR limits, we fix the threshold as described in
Section~\ref{subsec:practical_problem}.
The cyclic prefix (CP) is appended after PC to the final
waveforms of both the data and compensation antennas.}
\section{In-Band Distortion Compensation Using a Subset of Antennas for mMIMO-OFDM Systems}
\label{Sect:Prop}

This section formulates the in-band distortion compensation scheme by modeling the division of spatial resources as a partitioning problem. We consider the total set of $M$ antennas and partition them into a data-transmitting set and an in-band distortion compensation set.

The total set of antenna indices is denoted by $\mathcal{M} = \{1, 2, \ldots, M\}$. The core of the problem is the partitioning of $\mathcal{M}$ into two disjoint subsets: $\mathcal{M}_\mathrm{d}$ with cardinality $M_\mathrm{d}$ for transmitting data signals and $\mathcal{M}_\mathrm{e}$ with cardinality $M_\mathrm{e}$ for transmitting in-band distortion compensation signals, where $\mathcal{M} = \mathcal{M}_\mathrm{d} \cup \mathcal{M}_\mathrm{e}$.

This partitioning is represented by two selection matrices,
$\mathbf{S}_\mathrm{d} \in \{0,1\}^{M_\mathrm{d} \times M}$ for the $M_\mathrm{d}$ data antennas and 
$\mathbf{S}_\mathrm{e} \in \{0,1\}^{M_\mathrm{e} \times M}$ for the $M_\mathrm{e}$ in-band distortion compensation antennas.
These matrices consist of zeros and ones, satisfying $\mathbf{S}_\mathrm{d}\mathbf{S}_\mathrm{d}^\mathrm{T} = \mathbf{I}_{M_\mathrm{d}}$, $\mathbf{S}_\mathrm{e}\mathbf{S}_\mathrm{e}^\mathrm{T} = \mathbf{I}_{M_\mathrm{e}}$, and $\mathbf{S}_\mathrm{d}\mathbf{S}_\mathrm{e}^\mathrm{T} = \mathbf{0}$.

Using these matrices, the effective channels for the data transmission and in-band distortion compensation partitions are derived from the channel, given as
\begin{align}
  \mathbf{H}_\mathrm{d}[l]&= \mathbf{H}[l] \mathbf{S}_\mathrm{d}^\mathrm{T} \in \mathbb{C}^{U \times M_\mathrm{d}}, \nonumber\\
  \mathbf{H}_\mathrm{e}[l] &= \mathbf{H}[l] \mathbf{S}_\mathrm{e}^\mathrm{T} \in \mathbb{C}^{U \times M_\mathrm{e}}.
  \label{eq:branch_ch}
\end{align}
\red{In Fig.~\ref{fig:system_model},
$\widehat{\mathbf H}_{\mathrm d}$ denotes the data-branch
channel estimate, which equals $\mathbf H_{\mathrm d}$
under the assumed perfect channel state information.}
Digital precoders $\mathbf{W}_\mathrm{d}[l] \in \mathbb{C}^{M_\mathrm{d} \times U}$ and $\mathbf{W}_\mathrm{e}[l] \in \mathbb{C}^{M_\mathrm{e} \times U}$ are then designed based respectively on $\mathbf{H}_\mathrm{d}[l]$ and $\mathbf{H}_\mathrm{e}[l]$, typically using the RZF criterion from \eqref{eq:RZF_kappa}. 
The power normalization factor $\alpha_\mathrm d$ is chosen to satisfy the total transmission power constraint of $\mathbf{W}_\mathrm{d}[l]$. 
The power normalization factor $\alpha_\mathrm e$ is explained in~\eqref{eq:compNorm}.

The signal transmitted from the data antennas, after applying PC, is $\mathbf{x}'[l] \in \mathbb{C}^{M_\mathrm{d} \times 1}$, given as
\begin{equation}
  \mathbf{x}'[l] = \mathbf{W}_\mathrm{d}[l] \mathbf{s}[l] + \mathbf{d}[l],
\end{equation}
where $\mathbf{s}[l]\in\mathbb C^{U\times 1}$ is the user symbol vector and $\mathbf{d}[l]\in\mathbb C^{M_\mathrm d\times 1}$ is the in-band distortion signal.

To cancel the effect of this in-band distortion at the user terminals, an in-band distortion compensation signal $\mathbf{z}[l] \in \mathbb{C}^{M_\mathrm{e} \times 1}$ is generated and transmitted from the in-band distortion compensation RF chains:
\begin{equation}
  \mathbf{z}[l] = -\mathbf{W}_\mathrm{e}[l] \mathbf{H}_\mathrm{d}[l] \mathbf{d}[l].
\end{equation}
\redC{
The two signals are mapped to their assigned antennas by the selection
matrices, which gives the full-array digital PA-input vector
$\mathbf{v}[l]\in\mathbb{C}^{M\times 1}$ as
\begin{equation}
  \mathbf{v}[l] = \mathbf{S}_\mathrm{d}^\mathrm{T} \mathbf{x}'[l] + \mathbf{S}_\mathrm{e}^\mathrm{T} \mathbf{z}[l].
  \label{eq:v}
\end{equation}
The PAs act on the time-domain waveform corresponding to $\mathbf{v}[l]$.
Let $\bm{\epsilon}_\mathrm d[l]\in\mathbb{C}^{M_\mathrm{d}\times 1}$ and
$\bm{\epsilon}_\mathrm e[l]\in\mathbb{C}^{M_\mathrm{e}\times 1}$ denote the resulting PA-induced nonlinear
distortion at the data and compensation antennas, respectively.
The full-array PA-distortion vector $\bm{\epsilon}[l]\in\mathbb{C}^{M\times 1}$ is then
\begin{equation}
  \bm{\epsilon}[l] = \mathbf{S}_\mathrm{d}^\mathrm{T}\bm{\epsilon}_\mathrm d[l]
  + \mathbf{S}_\mathrm{e}^\mathrm{T}\bm{\epsilon}_\mathrm e[l].
  \label{eq:eps}
\end{equation}
By using \eqref{eq:branch_ch}-\eqref{eq:eps}, the received signal vector
$\mathbf{y}[l]\in\mathbb{C}^{U\times 1}$ is given by
\begin{align}
\mathbf{y}[l]
&= \mathbf{H}[l]\bigl(\mathbf{v}[l]+\bm{\epsilon}[l]\bigr)+\mathbf{n}[l] \nonumber\\
&= \mathbf{H}[l]\bigl(\mathbf{S}_\mathrm{d}^\mathrm{T}\mathbf{x}'[l]+\mathbf{S}_\mathrm{e}^\mathrm{T}\mathbf{z}[l]\bigr)
   +\mathbf{H}[l]\bigl(\mathbf{S}_\mathrm{d}^\mathrm{T}\bm{\epsilon}_\mathrm d[l]+\mathbf{S}_\mathrm{e}^\mathrm{T}\bm{\epsilon}_\mathrm e[l]\bigr)+\mathbf{n}[l] \nonumber\\
&= \mathbf{H}_\mathrm{d}[l]\mathbf{W}_\mathrm{d}[l]\mathbf{s}[l]
   +\bigl(\mathbf{I}_{U}-\mathbf{H}_\mathrm{e}[l]\mathbf{W}_\mathrm{e}[l]\bigr)
   \mathbf{H}_\mathrm{d}[l]\mathbf{d}[l] \nonumber\\
&\quad +\mathbf{H}_\mathrm{d}[l]\bm{\epsilon}_\mathrm d[l]
   +\mathbf{H}_\mathrm{e}[l]\bm{\epsilon}_\mathrm e[l]+\mathbf{n}[l] \nonumber\\
&= \mathbf{H}_\mathrm{d}[l]\mathbf{W}_\mathrm{d}[l]\mathbf{s}[l]
   +\mathbf r_\mathrm{comp}[l]+\bm{\nu}[l]+\mathbf{n}[l],
\label{eq:principle}
\end{align}
where $\bm{\nu}[l]\triangleq\mathbf{H}_\mathrm{d}[l]\bm{\epsilon}_\mathrm d[l]
+\mathbf{H}_\mathrm{e}[l]\bm{\epsilon}_\mathrm e[l]$
is the received PA distortion, $\mathbf{n}[l]\in\mathbb{C}^{U\times 1}$
is the noise vector, and
\begin{equation}
  \mathbf r_\mathrm{comp}[l]
  =
  (\mathbf I_U-\mathbf H_\mathrm e[l]\mathbf W_\mathrm e[l])\mathbf H_\mathrm d[l]\mathbf d[l]
  \label{eq:rcomp}
\end{equation}
is the compensation residual.}
Under RZF, this residual is generally nonzero.
It vanishes in the ZF special case with $\lambda_\mathrm e=0$ and
$\alpha_\mathrm e=1$. Hence, $\alpha_\mathrm e$ in \eqref{eq:Wkappa_norm} should ideally be unity so that the compensation signal reconstructs the user-domain PC distortion \(\mathbf H_\mathrm d[l]\mathbf d[l]\). 
The cancellation signal is generated to suppress peaks
exceeding the PC threshold and typically has a lower
average power than the data signal.
However, to prevent extreme cases, we set compensation power threshold $P_{\mathrm{cth}}$ to maintain system stability whose setting is specified in Section~\ref{subsec:practical_problem}, so that the compensation precoding matrix is given by
\begin{gather}
  \alpha_\mathrm e=\min\!\left\{1,\ \sqrt{\frac{P_{\mathrm{cth}}}{\sum_{l\in \mathcal{L}}\|\mathbf{W}_{\mathrm e,0}[l] \mathbf{H}_\mathrm{d}[l] \mathbf{d}[l]\|_\mathrm F^2}}\right\},\notag\\
  \mathbf{W}_{\mathrm e}[l]=\alpha_\mathrm e\,\mathbf{W}_{\mathrm e,0}[l],
\label{eq:compNorm}
\end{gather}
where the unnormalized precoding matrix $\mathbf{W}_{\mathrm e,0}[l]$ is calculated according to \eqref{eq:RZF_kappa}.

\redC{Since PC limits the data-waveform peaks, the data-branch PA distortion is
mitigated without distortion-aware data precoding, whereas the PA
distortion of the non-peak-limited compensation signal directly degrades
the cancellation and must be accounted for in the compensation
precoder~\cite{Li_VTC}. To optimize this precoder jointly with the antenna
partition without waveform simulation of every candidate, we next derive
an analytical SIDNR model that captures both PC and PA distortion.}

\section{SIDNR Analysis with Nonlinear Distortion and Compensation}
\label{Sect:SIDNR}

This section derives an analytical approximation of the SIDNR for the proposed PC based transmitter with in-band distortion compensation antennas and PA nonlinearity.

\subsection{Analytical Assumptions and Approximations}
\label{Sect:AAA}
\subsubsection{Bussgang-type decomposition}
For a memoryless nonlinearity acting on a Gaussian input, the output can be decomposed into a scaled version of the input plus an uncorrelated distortion term~\cite{Bussgang1952},
\begin{equation}
  \mathcal{Q}(x) = c\,x + \eta,\qquad \mathbb{E}[x^{\!*}\eta]=0,
  \label{eq:bussgang_generic}
\end{equation}
where the coefficient $c=\frac{\mathbb{E}[x^{\!*}\mathcal{Q}(x)]}{\mathbb{E}[|x|^2]}$ and $\eta$ represents the nonlinear distortion.

\subsubsection{Time-domain Gaussian approximation}
Conditioned on fixed precoders, we assume independent, zero-mean proper
symbols from a fixed finite constellation, with variance $\sigma_s^2>0$.
For data antenna $m\in\mathcal M_{\mathrm d}$, the inverse DFT (IDFT) of
$\mathbf x[l]=\mathbf W_{\mathrm d}[l]\mathbf s[l]$ is
\begin{equation}
  x_m[n]= \sum_{l\in\mathcal L}\sum_{u=1}^{U}a_{l,u}^{(m,n)}s_u[l],
  \label{eq:td_ofdm_sum}
\end{equation}
where $a_{l,u}^{(m,n)}\triangleq
w_{\mathrm d,m,u}[l]e^{j2\pi ln/N}/\sqrt N$.
If no weighted symbol dominates the variance, i.e.,
\begin{equation}
  \frac{\max_{l\in\mathcal L,u}|a_{l,u}^{(m,n)}|^2}
       {\sum_{l\in\mathcal L,u}|a_{l,u}^{(m,n)}|^2}
  \xrightarrow[|\mathcal L|\to\infty]{}0,
  \label{eq:lyapunov_condition}
\end{equation}
Lyapunov's central limit theorem~\cite{BillingsleyPM} supports the marginal approximation
   \(x_m[n]\sim\mathcal{CN}(0,\sigma_{x,m}^2)\),
   \redC{since, for a finite constellation, \eqref{eq:lyapunov_condition} implies
   Lyapunov's condition for the independent but non-identically distributed
   terms $a_{l,u}^{(m,n)}s_u[l]$.}
Independence across subcarriers gives
\begin{align}
  \sigma_{x,m}^2
  &\triangleq\mathbb E[|x_m[n]|^2]
  =\frac{1}{N}\sum_{l\in\mathcal L}\mathbb E[|x_m[l]|^2]\nonumber\\
  &=\frac{\sigma_s^2}{N}\sum_{l\in\mathcal L}
  [\mathbf W_{\mathrm d}[l]\mathbf W_{\mathrm d}^{\mathrm H}[l]]_{mm}.
  \label{eq:sig_xm_parseval}
\end{align}

For the PA input, we additionally adopt a Gaussian model $ v_m[n]\sim\mathcal{CN}(0,\sigma_{v,m}^2),$ with
symbol-averaged power,
\begin{equation}
  \sigma_{v,m}^2\triangleq
  \frac{1}{N}\sum_{t=0}^{N-1}\mathbb E[|v_m[t]|^2]
  =\frac{1}{N}\sum_{l\in\mathcal L}\mathbb E[|v_m[l]|^2].
  \label{eq:td_gaussian_v}
\end{equation}

\subsubsection{Distortion covariance approximation}
Let $\sigma_{\mathrm{pc},m}^2\triangleq\mathbb E[|d_m[n]|^2]$ and
$\sigma_{\mathrm{pa},m}^2\triangleq\mathbb E[|\epsilon_m[n]|^2]$ denote the
per-antenna time-domain distortion powers. 
Define the normalized input spectra as
\begin{align}
  \pi_{x,m}[l]
  &\triangleq
  \frac{\mathbb E[|x_m[l]|^2]}
       {\sum_{k\in\mathcal L}\mathbb E[|x_m[k]|^2]},\nonumber\\
  \pi_{v,m}[l]
  &\triangleq
  \frac{\mathbb E[|v_m[l]|^2]}
       {\sum_{k\in\mathcal L}\mathbb E[|v_m[k]|^2]}.
  \label{eq:pivm_def}
\end{align}
For $l\in\mathcal L$, we approximate the subcarrier-wise distortion powers by
\begin{align}
  \mathbb E[|d_m[l]|^2]
  &\approx N\sigma_{\mathrm{pc},m}^2\pi_{x,m}[l],
  \nonumber\\
  \mathbb E[|\epsilon_m[l]|^2]
  &\approx N\sigma_{\mathrm{pa},m}^2\pi_{v,m}[l].
  \label{eq:subcarrier_power_alloc}
\end{align}
These spectral proxies conserve the modeled total powers but are not implied
by Parseval's identity. The PA allocation additionally assigns all modeled
distortion power to the occupied band.

For the spatial structure, we note that per-antenna nonlinear processing
does not eliminate spatial distortion correlation~\cite{Hoffmann2026PA}. The
corresponding distortion covariances are approximated as
\begin{align}
  \mathbf C_{\mathbf d}[l]\triangleq\mathbb E[\mathbf d[l]\mathbf d^{\mathrm H}[l]]
  &\approx
  \mathbf D_{\mathrm{pc}}[l]
  \mathbf\Phi_{\mathrm{pc}}[l]
  \mathbf D_{\mathrm{pc}}[l],
  \nonumber\\
  \mathbf C_{\mathbf\epsilon,\kappa}[l]\triangleq
\mathbb E[\bm\epsilon_\kappa[l]\bm\epsilon_\kappa^{\mathrm H}[l]]
  &\approx
  \mathbf D_{\mathrm{pa},\kappa}[l]
  \mathbf\Phi_{\mathrm{pa},\kappa}[l]
  \mathbf D_{\mathrm{pa},\kappa}[l],
  \label{eq:Ceps_DPhiD}
\end{align}
where $\kappa\in\{\mathrm d,\mathrm e\}$; $\mathbf\Phi_{\mathrm{pc}}[l]$ and
$\mathbf\Phi_{\mathrm{pa},\kappa}[l]$ are unit diagonal correlation matrices.
From \eqref{eq:subcarrier_power_alloc}, 
for $m\in\mathcal M_{\mathrm d}$, $\left[\mathbf D_{\mathrm{pc}}[l]\right]_{mm}
=\sqrt{N\sigma_{\mathrm{pc},m}^2\pi_{x,m}[l]}$; for $m\in\mathcal M_\kappa$, $\left[\mathbf D_{\mathrm{pa},\kappa}[l]\right]_{mm}
=\sqrt{N\sigma_{\mathrm{pa},m}^2\pi_{v,m}[l]}$.
For a positive-semidefinite matrix $\mathbf Z$ with positive diagonal entries,
define
\begin{equation}
  \mathcal R(\mathbf Z)\triangleq
  \mathbf D_{\mathbf Z}^{-1/2}\mathbf Z\mathbf D_{\mathbf Z}^{-1/2},
  \qquad\mathbf D_{\mathbf Z}\triangleq\mathrm{diag}(\mathbf Z).
  \label{eq:spatial_normalization}
\end{equation}
Our precoder-guided correlation proxies are
\begin{align}
  \mathbf\Phi_{\mathrm{pc}}[l]
  &=\mathcal R\!\left(\mathbf W_{\mathrm d}[l]\mathbf\Upsilon_{\mathrm d}[l]
  \mathbf W_{\mathrm d}^{\mathrm H}[l]\right),
  \nonumber\\
  \mathbf\Phi_{\mathrm{pa},\kappa}[l]
  &=\mathcal R\!\left(\mathbf W_\kappa[l]\mathbf\Upsilon_{\mathrm{pa},\kappa}[l]
  \mathbf W_\kappa^{\mathrm H}[l]\right),
  \label{eq:Phi_pa_from_Wkappa}
\end{align}
where the $U\times U$ weighting matrices $\mathbf\Upsilon$ are positive
semidefinite. This construction preserves the allocated marginal powers.

For tractability, we neglect cross-correlation between the PA
distortions of the data and compensation branches $\mathbb E[\bm\epsilon_{\mathrm d}[l]\bm\epsilon_{\mathrm e}^{\mathrm H}[l]]$, while
retaining within-branch spatial correlations. This gives
\begin{equation}
  \mathbf C_{\mathbf\epsilon}[l]
  \approx\sum_{\kappa\in\{\mathrm d,\mathrm e\}}
  \mathbf S_\kappa^{\mathrm T}\mathbf C_{\mathbf\epsilon,\kappa}[l]\mathbf S_\kappa,
  \label{eq:Ceps_block}
\end{equation}
where $\mathbf C_{\mathbf\epsilon}[l]\triangleq
\mathbb E[\bm\epsilon[l]\bm\epsilon^{\mathrm H}[l]]$.
This cross-branch decorrelation is an additional approximation.

\subsection{PC-Induced In-Band Distortion Statistics}
For data antenna $m$, define the normalized threshold
\begin{equation}
  \gamma_m\triangleq\frac{A_{\mathrm{th}}}{\sigma_{x,m}},
  \label{eq:gamma_norm}
\end{equation}
where $\sigma_{x,m}^2$ is the average power of the pre-PC signal
at data antenna $m$, given by \eqref{eq:sig_xm_parseval}.

We adopt the level-crossing and sinc peak-shape approximations for
Gaussian OFDM signals in~\cite{SongOchiai2016}. 
The long-sinc limit, i.e., an untruncated sinc kernel, is used as an analytical approximation to our finite circular kernel.
All moments below
include averaging over one OFDM symbol.

The actual cancellation signal satisfies
\begin{equation}
  d_m[n]=(c_{\mathrm{pc},m}-1)x_m[n]+\eta_{\mathrm{pc},m}[n],
  \label{eq:pc_d_eta_relation}
\end{equation}
where $c_{\mathrm{pc},m}\triangleq
\mathbb E[x_m^*[n]x'_m[n]]/\sigma_{x,m}^2$ and
$\mathbb E[x_m^*[n]\eta_{\mathrm{pc},m}[n]]=0$.
Thus, $d_m[n]$ contains a signal-correlated component, unlike the
orthogonal residual $\eta_{\mathrm{pc},m}[n]$.

Following the analysis in \cite{SongOchiai2016} for the PC gain and cancellation power, with the latter evaluated in the long-sinc limit and scaled by $\sigma_{x,m}^2$, we obtain
\begin{align}
  c_{\mathrm{pc},m}
  &\approx 1-\frac{\mathrm{Si}(2\pi)}{\sqrt{3\pi}}
  \left(\gamma_m e^{-\gamma_m^2}
  +\sqrt{\pi}\,\mathrm{erfc}(\gamma_m)\right),
  \nonumber\\
  \sigma_{\mathrm{pc},m}^2
  &\triangleq\mathbb E[|d_m[n]|^2]
  \approx\frac{\pi\sigma_{x,m}^2}{2\sqrt{3}}
  \,\mathrm{erfc}(\gamma_m),
  \label{eq:alpha_songochiai}
\end{align}
where $\mathrm{Si}(\cdot)$ and $\mathrm{erfc}(\cdot)$ denote the sine
integral and complementary error function, respectively.
Since our kernel is confined to the occupied subcarriers,
$\sigma_{\mathrm{pc},m}^2$ is the in-band cancellation power used in
\eqref{eq:subcarrier_power_alloc} and \eqref{eq:Ceps_DPhiD}.
The orthogonal residual power follows as
\begin{equation}
  \sigma_{\eta_{\mathrm{pc}},m}^2
  =\sigma_{\mathrm{pc},m}^2
  -|c_{\mathrm{pc},m}-1|^2\sigma_{x,m}^2.
  \label{eq:sigma_eta_pc_from_power}
\end{equation}

With $\mathbf G_{\mathrm{pc}}\triangleq
\mathrm{diag}(c_{\mathrm{pc},1},\ldots,c_{\mathrm{pc},M_{\mathrm d}})$,
the frequency-domain decomposition is
\begin{equation}
  \mathbf x'[l]=\mathbf G_{\mathrm{pc}}\mathbf x[l]
  +\bm\eta_{\mathrm{pc}}[l],
  \label{eq:pc_bussgang_vector}
\end{equation}
where the covariance of $\bm\eta_{\mathrm{pc}}[l]$ is denoted by
$\mathbf C_{\eta_{\mathrm{pc}}}[l]$.

\subsection{PA-Induced Nonlinear Distortion Statistics}
Each active RF chain uses a memoryless Rapp solid-state power amplifier
(SSPA) with no AM/PM conversion~\cite{Rapp1991}. Its distortion statistics
are evaluated by following the Gaussian-input PA analysis in~\cite{Ochiai2013}.
The AM/AM characteristic of the SSPA is
\begin{equation}
  g(r)=Ar\left[1+\left(\frac{r}{r_{\max}}\right)^{2p}\right]^{-1/(2p)},
  \label{eq:sspa_rapp_short}
\end{equation}
where $r=|v_m[n]|$ is the PA-input envelope, $A$ is the small-signal
amplitude gain, $r_{\max}$ is the input saturation-reference amplitude,
and $p>0$ controls the transition smoothness. Without loss of generality, we set $A=1$.
The per-antenna input back-off (IBO) is
\begin{equation}
  \xi_m\triangleq\frac{r_{\max}^2}{\sigma_{v,m}^2},
  \label{eq:ibo_short}
\end{equation}
where $\sigma_{v,m}^2$ is the symbol-averaged PA-input power in
\eqref{eq:td_gaussian_v}.

Under the Gaussian approximation in
\eqref{eq:td_gaussian_v}, the input envelope is Rayleigh distributed~\cite{Ochiai2013},
yielding
\begin{align}
  \sigma_{\mathrm{pa},m}^2
  &\triangleq\mathbb E[|\epsilon_m[n]|^2]\nonumber\\
  &\approx\frac{2}{\sigma_{v,m}^2}
  \int_0^\infty r\,[g(r)-r]^2
  e^{-r^2/\sigma_{v,m}^2}\,dr.
  \label{eq:sigma_pa_direct}
\end{align}
This one-dimensional integral gives the total PA-error power used in
\eqref{eq:subcarrier_power_alloc}, where all modeled PA-error power
is assigned to the occupied band. The normalized power
$\sigma_{\mathrm{pa},m}^2/\sigma_{v,m}^2$ depends only on $(\xi_m,p)$
and can be tabulated offline.

\subsection{Per-User SIDNR Derivation}
\label{subsec:per_user_sidnr}
For a fixed channel realization, antenna partition, and applied
precoders, expectations are taken over the data symbols and receiver
noise. On subcarrier $l\in\mathcal L$, let
$\mathbf h_{\kappa,u}^{\mathrm T}[l]$ be row $u$ of
$\mathbf H_\kappa[l]$ for $\kappa\in\{\mathrm d,\mathrm e\}$,
and let $\mathbf w_{\mathrm d,j}[l]$ be column $j$ of
$\mathbf W_{\mathrm d}[l]$. Define
\begin{align}
  \beta_{u,j}[l]
  &\triangleq
  \mathbf h_{\mathrm d,u}^{\mathrm T}[l]\mathbf w_{\mathrm d,j}[l],\nonumber\\
  \mathbf g_u^{\mathrm T}[l]
  &\triangleq
  \mathbf h_{\mathrm d,u}^{\mathrm T}[l]
  -\mathbf h_{\mathrm e,u}^{\mathrm T}[l]
  \mathbf W_{\mathrm e}[l]\mathbf H_{\mathrm d}[l].
\label{eq:user_coefficients}
\end{align}
The $u$-th entry of \eqref{eq:principle} is then
\begin{equation}
  y_u[l]
  =\beta_{u,u}[l]s_u[l]
  +\sum_{j\ne u}\beta_{u,j}[l]s_j[l]+\mathbf g_u^{\mathrm T}[l]\mathbf d[l]
  +\nu_u[l]+n_u[l],
\label{eq:yu_compact}
\end{equation}
where $\nu_u[l]=\sum_{\kappa\in\{\mathrm d,\mathrm e\}}
\mathbf h_{\kappa,u}^{\mathrm T}[l]\bm\epsilon_\kappa[l]$.

We define the SIDNR relative to the nominal desired response
$\beta_{u,u}[l]s_u[l]$ as
\begin{equation}
  \Gamma_u[l]\triangleq
  \frac{\sigma_s^2|\beta_{u,u}[l]|^2}
  {\mathbb E[|y_u[l]-\beta_{u,u}[l]s_u[l]|^2]}.
  \label{eq:sidnr_reference}
\end{equation}
The average powers of the residual PC error after compensation
and the PA error at user $u$ are respectively given by
\begin{equation}
\begin{aligned}
  D_{\mathrm{pc},u}[l]
  &\triangleq
  \mathbb E\!\left[
    |\mathbf g_u^{\mathrm T}[l]\mathbf d[l]|^2
  \right]=\mathbf g_u^{\mathrm T}[l]
  \mathbf C_{\mathbf d}[l]\mathbf g_u^*[l],\\
  D_{\mathrm{pa},u}[l]
  &\triangleq
  \mathbb E[|\nu_u[l]|^2]\approx
  \sum_{\kappa\in\{\mathrm d,\mathrm e\}}
  \mathbf h_{\kappa,u}^{\mathrm T}[l]
  \mathbf C_{\mathbf\epsilon,\kappa}[l]
  \mathbf h_{\kappa,u}^*[l].
\end{aligned}
\label{eq:user_distortion_powers}
\end{equation}
The PA-power approximation follows from the cross-branch
uncorrelated model in \eqref{eq:Ceps_block}.

To evaluate \eqref{eq:sidnr_reference}, we approximate the
multiuser-interference term, the residual PC error after compensation,
and the PA error as zero-mean, pairwise uncorrelated contributions.
Their individual powers are retained and summed in the SIDNR
approximation.
This avoids additional joint-statistics modeling or waveform averaging
for each candidate configuration. Using the time-domain Gaussian and distortion-covariance approximations in Section~\ref{Sect:AAA} and independent zero-mean
receiver noise of variance $\sigma_n^2$, we obtain
\begin{equation}
  \Gamma_u[l]\approx
  \frac{\sigma_s^2|\beta_{u,u}[l]|^2}
  {\sigma_s^2\sum_{j\ne u}|\beta_{u,j}[l]|^2
  +D_{\mathrm{pc},u}[l]+D_{\mathrm{pa},u}[l]+\sigma_n^2}.
  \label{eq:sidnr_compact}
\end{equation}

\section{Proposed Rate Metric and Optimization Framework}
\label{Sect:Opt}

Based on the SIDNR model in Section~\ref{Sect:SIDNR}, we formulate the transmitter design in terms of a compact SIDNR-based rate metric and then specialize it to the operating regime considered in this paper.
To avoid repeated waveform generation for all candidates, we propose a two-stage procedure in which the candidates are first generated and ranked using the analytical SIDNR model, and waveform-based refinement is then applied only to a small shortlist.
Within the analytical stage, the optimization is separated into an inner compensation precoder design and an outer antenna-partition design.

\subsection{SIDNR-Based Average Spectral Efficiency and General Design Problem}
\label{subsec:general_problem}
For a fixed channel realization and fixed system parameters, we define the aggregate SIDNR-based rate metric as
\begin{equation}
R(\boldsymbol\theta)
\triangleq
\sum_{l\in\mathcal L}\sum_{u=1}^{U}
\log_2\!\left(1+\Gamma_u[l;\boldsymbol\theta]\right),
\label{eq:rate_metric}
\end{equation}
where the design vector $\boldsymbol\theta\triangleq(\mathcal M_{\mathrm e},A_{\mathrm{th}},\lambda_{\mathrm d},\lambda_{\mathrm e})$; \(\Gamma_u[l;\boldsymbol\theta]\) is the SIDNR evaluated from
\eqref{eq:sidnr_compact}.
Specifically, \(\mathcal M_{\mathrm e}\) determines the antenna
partition and hence the effective channels
\(\mathbf H_{\mathrm d}[l]\) and \(\mathbf H_{\mathrm e}[l]\);
\(A_{\mathrm{th}}\) determines the PC signal and its distortion
statistics; and \(\lambda_{\mathrm d}\) and
\(\lambda_{\mathrm e}\) determine the data- and compensation-side
precoders, respectively. 
\redC{The corresponding average spectral efficiency, normalized by the
numbers of users and occupied subcarriers, is defined as
\begin{equation}
\bar R(\boldsymbol\theta)
\triangleq
\frac{R(\boldsymbol\theta)}{U|\mathcal L|}
\quad \text{[bit/s/Hz/user]},
\label{eq:avg_se}
\end{equation}
which is also used as the performance metric in Section~\ref{Sect:PerfEval}.
Since $U|\mathcal L|$ is fixed, maximizing $\bar R(\boldsymbol\theta)$ is
equivalent to maximizing $R(\boldsymbol\theta)$.}

Let \(P_{\mathrm d}(\boldsymbol\theta)\) and
\(P_{\mathrm e}(\boldsymbol\theta)\) denote the corresponding PA-input powers of the data and compensation branches, respectively.
A general joint design can then be written as
\begin{subequations}
\begin{align}
\mathrm{(P0)}:\quad
\max_{\mathcal M_{\mathrm e},\,A_{\mathrm{th}},\,
      \lambda_{\mathrm d},\,\lambda_{\mathrm e}}
\quad & R(\boldsymbol\theta)\\
\mathrm{s.t.}\quad
&\mathcal M_{\mathrm d}=\mathcal M\setminus\mathcal M_{\mathrm e},\\
&A_{\mathrm{th}}^{\min}\le A_{\mathrm{th}}\le A_{\mathrm{th}}^{\max},\\
&\lambda_{\mathrm d}\ge0,\qquad\lambda_{\mathrm e}\ge0,\\
&P_{\mathrm d}(\boldsymbol\theta)
 +P_{\mathrm e}(\boldsymbol\theta)
 \le P_{\max}.
\end{align}
\end{subequations}
For a fixed PA saturation reference $r_{\max}$, the effective
per-antenna IBO in \eqref{eq:ibo_short} depends on
the design variables through the PA-input power:
\begin{equation}
  \xi_m(\boldsymbol\theta)
  =\frac{r_{\max}^2}
  {\sigma_{v,m}^2(\boldsymbol\theta)},
  \qquad
  \mathrm{IBO}_m(\boldsymbol\theta)
  =10\log_{10}\xi_m(\boldsymbol\theta)
  \quad \text{[dB]}.
  \label{eq:ibo_design_dependence}
\end{equation}
Here, $\sigma_{v,m}^2$ is evaluated after PC on the data branch
and after compensation precoding on the compensation branch,
in both cases before the PA.
Thus, the effective IBO is determined by $\boldsymbol\theta$,
and the corresponding PA-distortion statistics are included
in the SIDNR-based objective. 
Problem~\(\mathrm{(P0)}\) is nevertheless difficult
because the discrete partition changes both effective channels, while
\(A_{\mathrm{th}}\), \(\lambda_{\mathrm d}\), and \(\lambda_{\mathrm e}\)
jointly affect the PC residual, multiuser interference, and PA distortion.

\subsection{Practical Design Problem}
\label{subsec:practical_problem}
\redC{We next consider a practical operating regime of $\mathrm{(P0)}$
in which the PC threshold is fixed at $A_{\mathrm{th}}=A_0$,
chosen below and close to the PA saturation-reference amplitude,
which was shown to be effective in our previous study~\cite{Li_ACCESS}.}
This choice avoids an additional search over the PC threshold
and reduces the dimensionality and computational complexity
of the joint design, allowing the present work to focus on
antenna partitioning and compensation precoder optimization.
With this threshold setting, PC reduces the excursions of
the data waveform into the strongly nonlinear PA region.
The remaining data-branch PA distortion is consequently
treated as negligible in the design metric for the
operating regime considered here.
This approximation concerns only the additional distortion
introduced by the data-branch PA; the PC-induced in-band
distortion and its imperfect compensation are still accounted for.
The compensation waveform, in contrast, is generated through
the compensation precoder and is not constrained by the same
PC threshold.
Its peaks can therefore drive the compensation PAs into
nonlinear operation, so compensation-side PA distortion
must remain part of the precoder design~\cite{Li_VTC}.
Both branches retain their nonlinear PA models in the
waveform-level performance evaluation.

\red{Consistent with the above branch-specific design principle,
we adopt ZF data precoding, i.e., $\lambda_{\mathrm d}=0$,
to eliminate nominal multiuser interference and reduce the remaining design dimension.
For a full-row-rank $\mathbf H_{\mathrm d}[l]$, this yields
$\mathbf H_{\mathrm d}[l]\mathbf W_{\mathrm d}[l]
=\alpha_{\mathrm d}\mathbf I_U$.}
The pre-PC data signal is normalized to the power budget:
\begin{equation}
  \sum_{l\in\mathcal L}
  \mathrm{tr}\!\left(\mathbf C_x[l]\right)
  =P_{\max}.
  \label{eq:nominal_data_power}
\end{equation}
During candidate scoring, the
compensation precoder is evaluated with \(\alpha_{\mathrm e}=1\), so
excessive compensation power is reflected through the resulting PA
distortion rather than imposed as an explicit search constraint.
After operating-point selection, the compensation power is limited to
\(P_{\mathrm{cth}}=P_{\max}-P_{\mathrm d}\).
The final PA-input power therefore satisfies
\(
P_{\mathrm d}+P_{\mathrm e}
\le
P_{\max},
\)
which corresponds to the total-power budget in \(\mathrm{(P0)}\).
Thus, the practical design optimizes only the antenna partition and
the compensation-side regularization.
Consistent with the above
approximation, the practical design SIDNR is obtained from
\eqref{eq:sidnr_compact} by omitting only the data-branch PA-distortion
term:
\begin{equation}
\widetilde{\Gamma}_{u}[l]
\triangleq
\frac{\sigma_s^2|\beta_{u,u}[l]|^2}
{\sigma_s^2\sum_{j\ne u}|\beta_{u,j}[l]|^2
+D_{\mathrm{pc},u}[l]+D_{\mathrm{pa,e},u}[l]+\sigma_n^2},
\label{eq:sidnr_design}
\end{equation}
where
\(
D_{\mathrm{pa,e},u}[l]
=\mathbf h_{\mathrm e,u}^{\mathrm T}[l]
\mathbf C_{\mathbf\epsilon,\mathrm e}[l]
\mathbf h_{\mathrm e,u}^{*}[l]
\).

With $A_{\mathrm{th}}=A_0$ and $\lambda_{\mathrm d}=0$ fixed, the design vector
$\boldsymbol\theta$ reduces to $(\mathcal M_{\mathrm e},\lambda_{\mathrm e})$.
Replacing $\Gamma_u[l;\boldsymbol\theta]$ in \eqref{eq:rate_metric} with
$\widetilde\Gamma_u[l;\mathcal M_{\mathrm e},\lambda_{\mathrm e}]$, we obtain the
practical-design rate metric
\begin{equation}
\widetilde{R}(\mathcal M_{\mathrm e},\lambda_{\mathrm e})
\triangleq
\sum_{l\in\mathcal L}\sum_{u=1}^{U}
\log_2\!\left(1+\widetilde{\Gamma}_{u}[l;\mathcal M_{\mathrm e},\lambda_{\mathrm e}]\right),
\label{eq:rate_metric_design}
\end{equation}
where $\widetilde\Gamma_u[l;\mathcal M_{\mathrm e},\lambda_{\mathrm e}]$ is the SIDNR evaluated from
\eqref{eq:sidnr_design}.

Let \(\mathcal K\subseteq\{1,\ldots,M-U\}\) be the candidate set for the
number of compensation antennas. The practical problem solved in this
work is
\begin{subequations}
\begin{align}
\mathrm{(P1)}:\quad
\max_{\mathcal M_{\mathrm e},\,\lambda_{\mathrm e}}
\quad &\widetilde{R}(\mathcal M_{\mathrm e},\lambda_{\mathrm e})
\label{eq:P1_obj}\\
\mathrm{s.t.}\quad
&\mathcal M_{\mathrm d}=\mathcal M\setminus\mathcal M_{\mathrm e},
\label{eq:P1_partition}\\
&|\mathcal M_{\mathrm e}|\in\mathcal K,
\label{eq:P1_K}\\
&\mathrm{rank}\!\left(\mathbf H_{\mathrm d}[l]\right)=U,
\qquad \forall l\in\mathcal L,
\label{eq:P1_rank}\\
&\lambda_{\mathrm e}\ge 0.
\label{eq:P1_lambda}
\end{align}
\end{subequations}

\subsection{Inner-Layer Compensation Precoder Optimization}
\label{subsect:inner_algorithm}

For a fixed hard antenna partition, \(\mathrm{(P1)}\) reduces to the
scalar optimization
\begin{equation}
\lambda_{\mathrm e}^{\star}(\mathcal M_{\mathrm e})
=
\arg\max_{\lambda_{\mathrm e}}
\widetilde{R}(\mathcal M_{\mathrm e},\lambda_{\mathrm e}).
\label{eq:lambda_inner}
\end{equation}
The unnormalized compensation precoder is the scalar-RZF form in
\eqref{eq:RZF_kappa}. Small \(\lambda_{\mathrm e}\) values approach ZF
and suppress the user-domain PC residual aggressively, but can create a
large compensation waveform and hence strong compensation-side PA
distortion. Increasing \(\lambda_{\mathrm e}\) reduces this PA burden
at the cost of a larger residual PC term. Since only one scalar is
optimized, \eqref{eq:lambda_inner} is solved by a logarithmic grid
search.
Algorithm~\ref{alg:inner_design} summarizes the inner-layer method.

\begin{algorithm}[tb]
\caption{Inner-layer compensation precoder optimization}
\label{alg:inner_design}
\begin{algorithmic}[1]
\REQUIRE Hard partition \(\mathcal M_{\mathrm e}\), fixed threshold
\(A_0\), total power budget $P_{\max}$, and candidate grid
\(\mathcal G_{\lambda}\).
\ENSURE \(\lambda_{\mathrm e}^{\star}\) and the final compensation
precoder.
\STATE Construct \(\mathbf H_{\mathrm d}[l]\) and
\(\mathbf H_{\mathrm e}[l]\); compute the data-side ZF precoder and
normalize it according to \eqref{eq:nominal_data_power}.
\STATE Evaluate the analytical PC-distortion statistics for the fixed
threshold \(A_0\).
\FOR{each \(\lambda_{\mathrm e}\in\mathcal G_{\lambda}\)}
  \STATE Construct
  \(\mathbf W_{\mathrm e,0}[l;\lambda_{\mathrm e}]\) from
  \eqref{eq:RZF_kappa} and set \(\alpha_{\mathrm e}=1\) for candidate
  scoring.
  \STATE Evaluate the uncapped compensation-signal statistics and the
  corresponding compensation-side PA distortion.
  \STATE Evaluate \(\widetilde{\Gamma}_{u}[l]\) from
  \eqref{eq:sidnr_design} and compute
  \(\widetilde{R}(\mathcal M_{\mathrm e},\lambda_{\mathrm e})\).
\ENDFOR
\STATE Select \(\lambda_{\mathrm e}^{\star}\) according to
\eqref{eq:lambda_inner}.
\STATE If this partition is selected as the final transmitter
configuration, compute \(\alpha_{\mathrm e}\) from
\eqref{eq:compNorm}.
\STATE Set
\(
\mathbf W_{\mathrm e}^{\star}[l]
=
\alpha_{\mathrm e}
\mathbf W_{\mathrm e,0}[l;\lambda_{\mathrm e}^{\star}].
\)
\RETURN \(\lambda_{\mathrm e}^{\star}\) and
\(\{\mathbf W_{\mathrm e}^{\star}[l]\}_{l\in\mathcal L}\).
\end{algorithmic}
\end{algorithm}

\subsection{Outer-Layer Antenna-Partition Optimization}
\label{subsect:outer_algorithm}

The remaining difficulty in \(\mathrm{(P1)}\) is the combinatorial
antenna partition. To solve this, we introduce binary role variables
\(q_m\in\{0,1\}\), where \(q_m=0\) and \(q_m=1\) denote data and
compensation roles, respectively, and relax them to \(0\le q_m\le1\).
The relaxed variables define a virtual fractional RF-resource model used during optimization: \(q_m\) assigns the data and compensation sides
of antenna \(m\) the amplitude weights \(\sqrt{1-q_m}\) and \(\sqrt{q_m}\), i.e.,
\begin{align}
\mathbf A_{\mathrm d}(\mathbf q)
&=\mathrm{diag}\!\left(\sqrt{1-q_1},\ldots,\sqrt{1-q_M}\right),\nonumber\\
\mathbf A_{\mathrm e}(\mathbf q)
&=\mathrm{diag}\!\left(\sqrt{q_1},\ldots,\sqrt{q_M}\right),
\label{eq:Ade_relaxed}
\end{align}
and the relaxed channels are
\(\mathbf H_{\kappa}^{\mathrm r}[l;\mathbf q]=\mathbf H[l]\mathbf A_{\kappa}(\mathbf q)\),
\(\kappa\in\{\mathrm d,\mathrm e\}\). To preserve the amplitude
interpretation, the PC threshold and the PA saturation reference of
antenna \(m\) are scaled as
\begin{equation}
A_{0,m}^{\mathrm r}=\sqrt{1-q_m}\,A_0,\qquad
r_{\max,\mathrm e,m}^{\mathrm r}=\sqrt{q_m}\,r_{\max},
\label{eq:scaling_relaxed}
\end{equation}
so that the normalized threshold and the compensation-side IBO become
\(\gamma_m^{\mathrm r}=A_{0,m}^{\mathrm r}/\sigma_{x,m}^{\mathrm r}\) and
\(\xi_{\mathrm e,m}^{\mathrm r}=q_m r_{\max}^2/(\sigma_{v,\mathrm e,m}^{\mathrm r})^2\),
where \((\sigma_{x,m}^{\mathrm r})^2\) and \((\sigma_{v,\mathrm e,m}^{\mathrm r})^2\)
are the pre-PC and compensation-side PA-input powers obtained with the
relaxed precoders.
A side with zero weight, i.e., the data side for \(q_m=1\) or the compensation side for \(q_m=0\), carries no signal, so its zero threshold or saturation reference has no physical meaning.
All other quantities, i.e., the ZF precoder, the uncapped RZF compensation
precoder, the compensation signal, the PC and
compensation-side PA covariances, \(\beta_{u,j}[l]\), \(\mathbf g_u[l]\),
and the SIDNR~\eqref{eq:sidnr_design}, are computed as in the
hard-partition case with \(\mathbf H_{\kappa}[l]\) replaced by
\(\mathbf H_{\kappa}^{\mathrm r}[l;\mathbf q]\); we denote the resulting
relaxed SIDNR by \(\widetilde\Gamma_u^{\mathrm r}[l]\). The relaxed design metric
\(\widetilde R^{\mathrm r}(\mathbf q,\lambda_{\mathrm e})\) is then
\eqref{eq:rate_metric_design} with \(\widetilde\Gamma_u[l;\mathcal M_{\mathrm e},\lambda_{\mathrm e}]\) replaced by
\(\widetilde\Gamma_u^{\mathrm r}[l;\mathbf q,\lambda_{\mathrm e}]\).

To promote binary solutions, define
\begin{equation}
B(\mathbf q)\triangleq\sum_{m=1}^{M}q_m(1-q_m),
\label{eq:binary_penalty}
\end{equation}
which is zero for a binary vector.
The compensation-antenna count is selected through a
one-dimensional discrete search over $K\in\mathcal K$.
For each candidate $K$, the antenna placement is optimized
separately over
\begin{equation}
\mathcal Q_K\triangleq
\left\{\mathbf q:\;
0\le q_m\le1,\ \forall m,\quad
\mathbf 1^{\mathrm T}\mathbf q=K
\right\}.
\label{eq:relaxed_Q}
\end{equation}
We use the penalized continuous surrogate
\begin{equation}
F_{\mu}(\mathbf q,\lambda_{\mathrm e})
\triangleq
\widetilde{R}^{\mathrm r}(\mathbf q,\lambda_{\mathrm e})
-\mu B(\mathbf q),
\qquad
\mathbf q\in\mathcal Q_K.
\label{eq:relaxed_objective}
\end{equation}
This surrogate
remains nonconvex because the relaxed rate depends on \(\mathbf q\). We therefore alternate between
\(\lambda_{\mathrm e}\) and \(\mathbf q\).
For fixed \(\mathbf q\), the scalar regularization is updated by
\begin{equation}
\lambda_{\mathrm e}^{(t+1)}
=
\arg\max_{\lambda_{\mathrm e}}
\widetilde{R}^{\mathrm r}
(\mathbf q^{(t)},\lambda_{\mathrm e}).
\label{eq:lambda_line_search}
\end{equation}
For fixed \(\lambda_{\mathrm e}\), we construct a local minorizing surrogate at
\(\mathbf q^{(t)}\):
\begin{align}
\widetilde F_{\mu}(\mathbf q\mid\mathbf q^{(t)})
&=
\widetilde{R}^{\mathrm r}
(\mathbf q^{(t)},\lambda_{\mathrm e})
+
\nabla_{\mathbf q}\widetilde{R}^{\mathrm r}
(\mathbf q^{(t)},\lambda_{\mathrm e})^{\mathrm T}
(\mathbf q-\mathbf q^{(t)})
\nonumber\\
&\!\!\!\!\!\!\!\!\!\!\!\!\!\!\!\!
-\frac{L_q}{2}\|\mathbf q-\mathbf q^{(t)}\|_2^2
-\mu\!\left[
B(\mathbf q^{(t)})
+
\nabla_{\mathbf q}B(\mathbf q^{(t)})^{\mathrm T}
(\mathbf q-\mathbf q^{(t)})
\right],
\label{eq:q_surrogate_new}
\end{align}
where \(L_q>0\) is a proximal parameter. 
Since \(B(\mathbf q)\) is concave on \([0,1]^M\), its first-order approximation is a global upper bound, and hence the negative linearized penalty lower-bounds \(-\mu B(\mathbf q)\). 
Moreover, if \(L_q\) is no smaller than a local Lipschitz constant of \(\nabla_{\mathbf q}\widetilde R^{\mathrm r}\), the quadratic first-order model locally lower-bounds the relaxed rate. Intuitively, a larger \(L_q\) makes the approximation more conservative by reducing the update step.
Therefore, \(\widetilde F_\mu\) is a local minorizer of \(F_\mu\).

Maximizing \eqref{eq:q_surrogate_new} over \(\mathcal Q_K\) gives the
projected-gradient update
\begin{equation}
\mathbf q^{(t+1)}
=
\Pi_{\mathcal Q_K}\!\left[
\mathbf q^{(t)}
+
\frac{1}{L_q}
\left(
\nabla_{\mathbf q}\widetilde{R}^{\mathrm r}
(\mathbf q^{(t)},\lambda_{\mathrm e})
-\mu\nabla_{\mathbf q}B(\mathbf q^{(t)})
\right)
\right],
\label{eq:q_projected_update}
\end{equation}
where
\(\nabla_{\mathbf q}B(\mathbf q)
=[1-2q_1,\ldots,1-2q_M]^{\mathrm T}\) and 
\(
\Pi_{\mathcal Q_K}(\mathbf x)
\triangleq
\arg\min_{\mathbf q\in\mathcal Q_K}
\|\mathbf q-\mathbf x\|_2^2
\)
denotes the Euclidean projection onto \(\mathcal Q_K\).
In practice, a trial \(\mathbf q\)-update is accepted only when it does not decrease \(F_\mu\); otherwise, \(L_q\) is increased by backtracking and the projected update is recomputed. This ensures a nondecreasing \(F_\mu\) during the \(\mathbf q\)-update.

To avoid forcing a poor binary assignment too early, the penalty is
introduced by continuation:
\begin{equation}
\mu^{(r+1)}=\rho_{\mu}\mu^{(r)},\qquad \rho_{\mu}>1.
\label{eq:mu_continuation}
\end{equation}
After the relaxed search for a given $K$ terminates,
its final weight vector, denoted by $\mathbf q^\star(K)$,
is sorted in descending order.
The $K$ largest entries define the corresponding hard
candidate set $\mathcal M_{\mathrm e}^{(K)}$.
The scalar $\lambda_{\mathrm e}$ is then re-optimized
for this partition by Algorithm~\ref{alg:inner_design}
using the uncapped design metric, which gives
$\lambda_{\mathrm e}^{\star}(K)\triangleq\lambda_{\mathrm e}^{\star}(\mathcal M_{\mathrm e}^{(K)})$.
This procedure is repeated separately for every
$K\in\mathcal K$.

The hard candidates are first ranked in descending order
of their analytically optimized rate metrics
\(
\widetilde R\!\left(
\mathcal M_{\mathrm e}^{(K)},
\lambda_{\mathrm e}^{\star}(K)
\right)\) for
\( K\in\mathcal K.
\)
A small shortlist of the highest-scoring candidates,
indexed by $\mathcal K_{\mathrm s}\subseteq\mathcal K$,
is retained for waveform-based refinement.

For each shortlisted partition, the PC waveform is
explicitly generated, and the scalar search over
$\lambda_{\mathrm e}$ is repeated using the same
SIDNR-based rate expression, but with the analytical
PC and compensation-branch PA distortion covariances
replaced by their waveform-based sample counterparts.
Denote this metric by $\widetilde R_{\mathrm{wf}}$
and the corresponding optimized regularization parameter
by $\lambda_{\mathrm e,\mathrm{wf}}^\star(K)$.
The final compensation-antenna count is selected as
\begin{equation}
K^\star
=
\arg\max_{K\in\mathcal K_{\mathrm s}}
\widetilde R_{\mathrm{wf}}\!\left(
\mathcal M_{\mathrm e}^{(K)},
\lambda_{\mathrm e,\mathrm{wf}}^\star(K)
\right),
\label{eq:hard_recovery}
\end{equation}
with the final partition given by
$\mathcal M_{\mathrm e}^{\star}
=\mathcal M_{\mathrm e}^{(K^\star)}$.
It is worth mentioning that candidate-specific PC-waveform generation is restricted to $\mathcal K_{\mathrm s}$; the preceding candidate-generation and analytical screening stages do not require it. The compensation-power limiter in \eqref{eq:compNorm} is applied only
after this hard operating point has been selected.

\begin{algorithm}[t]
\caption{Outer-layer relaxed antenna-partition optimization}
\label{alg:relaxed_role}
\begin{algorithmic}[1]
\FOR{each \(K\in\mathcal K\)}
  \STATE Initialize \(\mathbf q^{(0)}\in\mathcal Q_K\),
  \(\lambda_{\mathrm e}^{(0)}\), and \(\mu=\mu_0\).
  \FOR{\(r=0,1,\ldots,R_{\max}\)}
    \REPEAT
      \STATE Update \(\lambda_{\mathrm e}\) by
      \eqref{eq:lambda_line_search}.
      \STATE Update \(\mathbf q\) by
      \eqref{eq:q_projected_update} over \(\mathcal Q_K\).
    \UNTIL{convergence or the iteration limit
    is reached.}
    \STATE Set \(\mu\leftarrow\rho_{\mu}\mu\).
  \ENDFOR
  \STATE Denote the final relaxed vector by
  \(\mathbf q^\star(K)\) and sort its entries
  in descending order.
  \STATE Form \(\mathcal M_{\mathrm e}^{(K)}\)
  from the \(K\) largest entries.
  \STATE Re-optimize \(\lambda_{\mathrm e}^{\star}(K)\)
  using Algorithm~\ref{alg:inner_design} and record
  the uncapped hard-partition metric \(\widetilde{R}\).
\ENDFOR
\STATE \red{Refine the analytically shortlisted candidates
using waveform-based sample distortion covariances,
select $K^\star$ by \eqref{eq:hard_recovery},
and apply \eqref{eq:compNorm} to the selected
refined compensation precoder.}
\end{algorithmic}
\end{algorithm}

\section{Performance Evaluation and Discussion}
\label{Sect:PerfEval}

\subsection{Simulation Setup}
\label{subsec:simulation_setup}

The BS employs a $10\times10$ UPA, comprising
$M=100$ antennas, and serves $U=10$ single-antenna users. The horizontal and vertical element spacings
are both half a wavelength. Each OFDM symbol uses a 512-point
IDFT/DFT with 64 occupied subcarriers, corresponding to an
oversampling factor of eight. The CP contains 128
samples, and 64-QAM is used on all occupied subcarriers.
Table~\ref{tab:simulation_parameters} summarizes the parameters.

A three-dimensional frequency-selective channel with six
equal-power non-line-of-sight (NLOS) delay taps is considered. Each tap is synthesized
from 16 unresolved rays, with a tap spacing of eight samples.
\red{Perfect channel state information is assumed.}
Horizontal and vertical ray offsets are independently and uniformly
distributed within $\pm5^{\circ}$ around the corresponding user
direction. 
The users are located on a circle around the BS, with independently
and uniformly distributed azimuth angles. The same user positions
are used across the compared schemes within each trial.
Distance-dependent large-scale attenuation is not considered.
One channel snapshot is independently generated per trial and
remains fixed over a block of 20 OFDM symbols. Channel realizations,
information bits, and receiver-noise random streams are paired
across schemes within each comparison experiment.

The PC threshold is 3~dB relative to the data signal power
reference. 
A full-symbol circular cancellation kernel of 512 samples is used, with at most 200 PC operations per data antenna and OFDM symbol. Each active antenna is followed by a memoryless
Rapp SSPA with smoothness parameter $p=200$ and no AM/PM conversion.
The nominal IBO is 4~dB relative to the common reference
$P_{\mathrm{avr}}=1$. The actual input power, and therefore the
effective back-off, can differ between antennas.

The main BER, received-SIDNR, and spectral-efficiency comparison uses
200 channel trials per $E_{\mathrm{s}}/N_0$ value,
where $E_{\mathrm{s}}/N_0$ denotes the ratio of the normalized average
transmit energy per data symbol of each user and subcarrier to the noise
power spectral density at each user.
Each trial contains 20 OFDM symbols.

\begin{table}[t]
  \centering
  \caption{Principal Simulation Parameters}
  \label{tab:simulation_parameters}
  \footnotesize
  \setlength{\tabcolsep}{3pt}
  \renewcommand{\arraystretch}{1.08}
  \begin{tabular}{|p{0.47\columnwidth}|p{0.43\columnwidth}|}
    \hline
    Parameter & Value \\
    \hline
    BS array & $10\times10$ UPA \\
    Number of BS antennas / users & 100 / 10 \\
    Horizontal/vertical spacing & $0.5\lambda$ \\
    DFT size / occupied subcarriers & 512 / 64 \\
    Oversampling factor / CP length & 8 / 128 samples \\
    Modulation & 64-QAM \\
    Channel & 3-D frequency-selective NLOS \\
    Delay taps / rays per tap & 6 / 16 \\
    Horizontal/vertical ray offsets & Uniform within $\pm5^{\circ}$ \\
    Channel evolution within a trial & Static \\
    PC threshold / pulse length & 3 dB / $L_\mathrm{pc} = 512$ samples \\
    Maximum PC operations & 200 per antenna and symbol \\
    PA model / smoothness parameter & Rapp SSPA / $p=200$ \\
    Nominal PA IBO & 4 dB \\
    Refinement shortlist size & $N_\mathrm s=5$ \\
    \hline
  \end{tabular}
\end{table}

\subsection{Compared Schemes and Evaluation Metrics}

The names used in the figures have the following meanings.
\emph{Random Partition} reserves $K=30$ randomly selected antennas
for distortion compensation.

\emph{Outer Placement} follows the placement design in~\cite{Li_ACCESS}.
Closely spaced compensation antennas have strongly correlated channels, so the ZF
precoder produces a strong, peaky compensation signal and
hence severe compensation-branch PA distortion. Assigning the compensation antennas
to the outer elements of the UPA maximizes their mutual distance and reduces this
correlation, which is beneficial when the PC-induced distortion is
significant. Here, $K=30$ compensation antennas are fixed on the
outermost ring with ZF compensation precoding, providing a reference for the
proposed adaptive partitioning.

\emph{Iterative Compensation} uses the entire antenna array and repeatedly compensates the channel-propagated PC distortion through frequency-domain precoding and time-domain waveform updates, thereby hiding the peaks in the channel null space. PC is reapplied between compensation rounds; the final compensated waveform is transmitted without another PC step. The main performance comparison uses 11 rounds. The separate runtime experiment considers
3, 7, 11, 20, and 30 rounds.

\emph{Proposed $\lambda$ Sweep} retains the fixed $K=30$ partition and optimizes $\lambda_{\mathrm e}$ through line search. 

\emph{Proposed Full} searches both the partition
and the compensation regularization using analytical PC/PA statistics and predicted SIDNR. 
\red{The compensation-antenna count is selected through a
one-dimensional discrete search over
$\mathcal K=\{25,26,\ldots,35\}$.}
The candidates with the highest scores are refined using the sample distortion covariances, and the output of this restricted search becomes the final partition.

For each trial, the waveform-measured user-subcarrier SIDNR is
\begin{equation}
\red{
\widehat{\Gamma}_u[l]
=
\frac{\sum_b |s_{u,b}[l]|^2}
{\sum_b |\widehat{s}_{u,b}[l]-s_{u,b}[l]|^2},
}
\label{eq:eval_measured_sidnr}
\end{equation}
where $b$ indexes the OFDM symbols within each channel trial,
and $s_{u,b}[l]$ and $\widehat{s}_{u,b}[l]$ denote the
ideal symbols and the decision points after receive processing,
respectively.
The residual includes PC and PA distortion, multiuser
interference, and noise.
For the per-user SIDNR cumulative distribution function (CDF), the numerator and denominator
are additionally summed over occupied subcarriers before
taking the ratio. The resulting user values are collected
over channel trials. 

The measured average spectral efficiency, i.e., \eqref{eq:avg_se} evaluated with $\widehat{\Gamma}_u[l]$, is
\begin{equation}
\bar R^{\mathrm{meas}}
=
\frac{1}{U|\mathcal L|}
\sum_{u=1}^{U}\sum_{l\in\mathcal L}
\log_2\!\left(1+\widehat{\Gamma}_u[l]\right).
\label{eq:measured_rate_per_user}
\end{equation}

\subsection{Simulation Results and Discussion}
\begin{figure}[t]
  \centering
  \includegraphics[width=\linewidth]{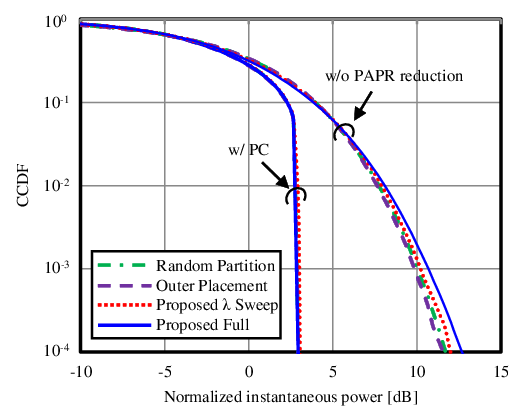}
  \caption{CCDFs of normalized instantaneous power of data-antenna signals before and after PC with a 3-dB threshold.}
  \label{fig:ccdf_data}
\end{figure}

\begin{figure}[t]
  \centering
  \includegraphics[width=\linewidth]{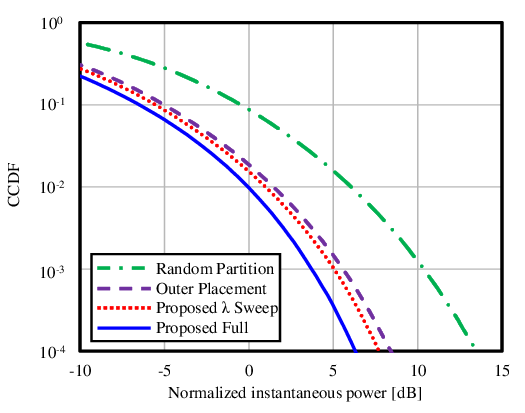}
  \caption{CCDFs of normalized instantaneous power of compensation-antenna signals, normalized by the same data signal power reference as in Fig.~\ref{fig:ccdf_data}.}
  \label{fig:ccdf_comp}
\end{figure}
\redC{To characterize the large-amplitude behavior of the
transmitted signals, we use the complementary cumulative
distribution function (CCDF), defined as
\(
\mathrm{CCDF}(x)=\Pr(X>x),
\)
where $X$ denotes the normalized instantaneous power
and $x$ is the corresponding power threshold.
The CCDF is used to compare the power distributions
of the data and compensation branches.}

Figure~\ref{fig:ccdf_data} shows a pronounced reduction in the
large-amplitude tail after PC. The post-PC curves decrease sharply
near the configured threshold, with only small differences between
the partition-based schemes. Consequently, differences in received
performance cannot be explained by data branch peak
suppression alone. 

The compensation branch behaves differently in
Fig.~\ref{fig:ccdf_comp}. Random Partition has a substantially
heavier upper tail than Outer Placement and the two proposed designs.
At a CCDF near $10^{-3}$, its normalized power is approximately
10~dB, compared with roughly 5~dB for the structured partitions.
This is consistent with compensation power being sensitive to the
selected channel submatrix and compensation precoder. The curve for Proposed $\lambda$ Sweep is slightly higher than that for Proposed Full.

\begin{figure}[t]
  \centering
  \includegraphics[width=\linewidth]{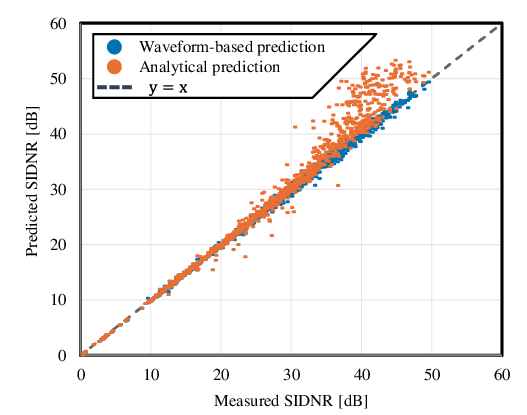}
  \caption{Predicted versus measured user-subcarrier SIDNR
        for Proposed Full. Blue points use waveform-based sample PC/PA distortion covariances;
        orange points use analytical correlated PC/PA
        covariances. The dashed line denotes $y=x$.}
  \label{fig:sidnr_prediction}
\end{figure}

Figure~\ref{fig:sidnr_prediction} separates two aspects of the SIDNR model. Both sets of points use waveform-measured SIDNR on the horizontal axis. The blue points evaluate the covariance-based expression
using waveform-based sample distortion covariances for the PC and compensation-branch PA components, whereas the orange points use their analytical approximations. 
The analytical predictions track the measured values well in the low-to-moderate SIDNR region, but exhibit increasing scatter and an optimistic bias for some high-SIDNR records.  This limitation becomes particularly relevant under low-noise conditions, where effective PC compensation can make the remaining PA distortion a dominant impairment.
This shows the limitation of the current Gaussian-input approximation used in the analytical PA model. The compensation waveform is generated from peak-dependent PC signals and spatial precoding, and its amplitude distribution need not follow the assumed Gaussian model. In particular, pronounced amplitude tails may not be adequately represented by the approximation. For the near-soft-limiting PA considered here, underestimating excursions into saturation can underestimate the resulting distortion power and consequently overestimate the SIDNR.

\begin{figure}[t]
  \centering
  \includegraphics[width=\linewidth]{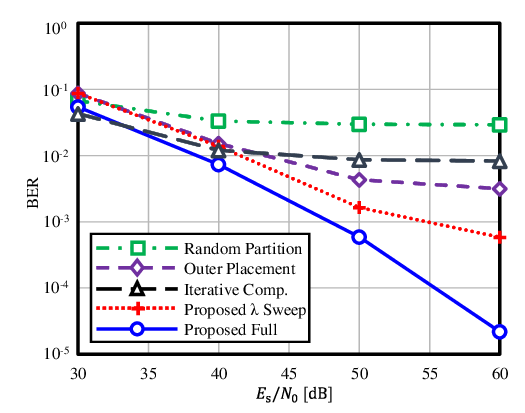}
  \caption{64-QAM BER versus
        $E_{\mathrm{s}}/N_0$ for the compared schemes.
        Iterative compensation uses 11 iterations;
        Proposed $\lambda$ Sweep uses a fixed $K=30$
        partition; and Proposed Full selects $K$
        from 25 to 35.}
  \label{fig:ber_comparison}
\end{figure}
Figure~\ref{fig:ber_comparison} shows that iterative compensation
has the lowest BER among the displayed schemes at 30~dB. 
At higher $E_{\mathrm{s}}/N_0$, the benefit of compensation precoder and partition
optimization becomes clearer. At 60~dB, the BER is approximately
$2\times10^{-5}$ for Proposed Full, compared with approximately
$6\times10^{-4}$ for Proposed $\lambda$ Sweep, $3\times10^{-3}$ for
Outer Placement, and $8\times10^{-3}$ for 11-round iterative
compensation. Proposed $\lambda$ Sweep improves upon the fixed-placement
baseline through regularization optimization. The further
improvement of Proposed Full demonstrates the additional benefit
of adapting antenna roles and their number under the tested
configuration.

\begin{figure}[t]
  \centering
  \includegraphics[width=\linewidth]{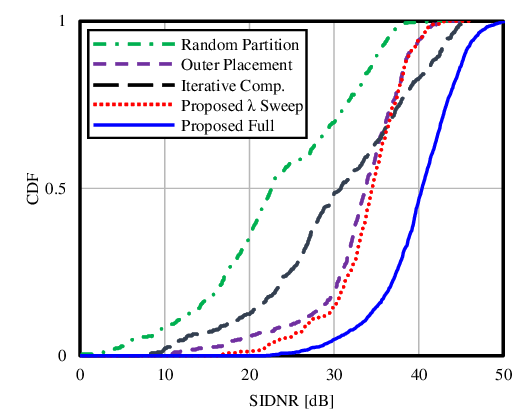}
  \caption{Empirical CDFs of waveform-measured
        per-user SIDNR at $E_{\mathrm{s}}/N_0=60$~dB.
        Each user value aggregates signal and residual
        powers over occupied subcarriers.}
  \label{fig:sidnr_cdf}
\end{figure}
The distribution in Fig.~\ref{fig:sidnr_cdf} complements the BER
comparison. Proposed Full shifts the user-level SIDNR distribution
toward higher values relative to Random Partition, Outer Placement,
and Proposed $\lambda$ Sweep. The median is near 40~dB for Proposed Full,
compared with approximately 34~dB for the two structured
fixed-partition schemes. This indicates that the improvement is
visible across the user distribution, rather than only in a single
average. 

\begin{figure*}[t]
  \centering
  \includegraphics[width=0.86\textwidth]{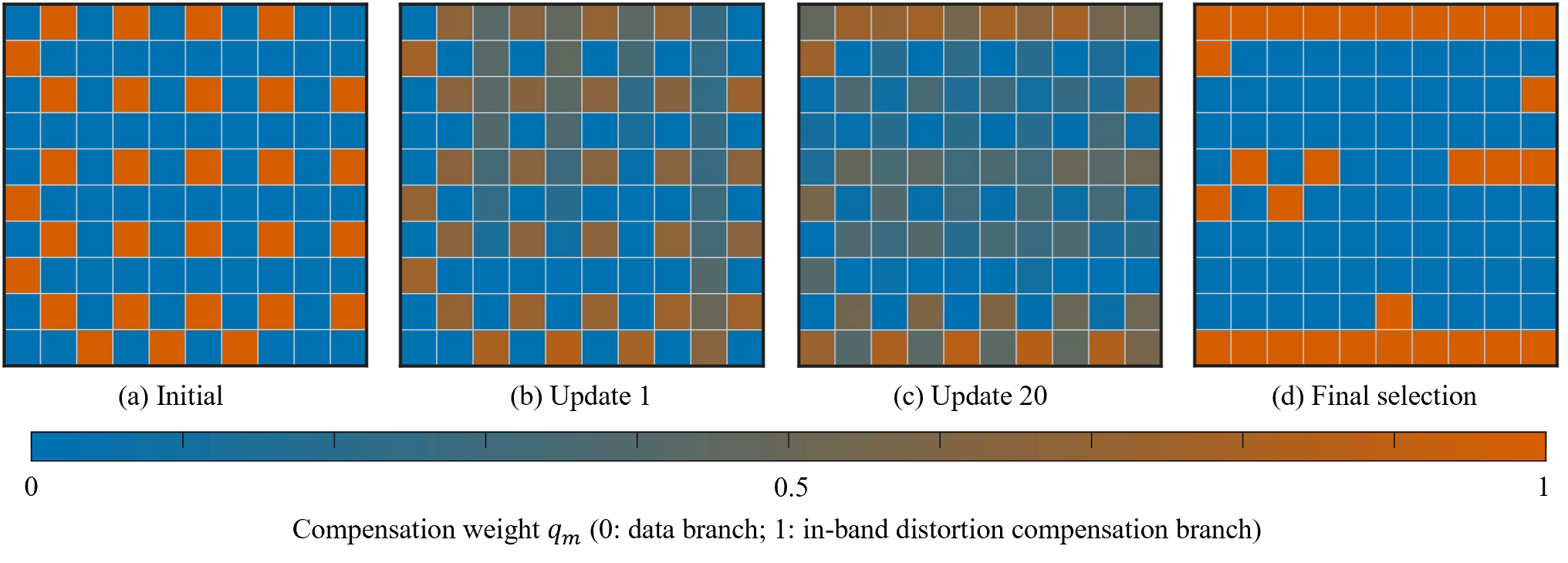}
\caption{Illustrative evolution of the antenna partition at $E_{\mathrm{s}}/N_0=60$~dB.
Panels (a)-(c) show the actual relaxed weights at updates
$t=0,1,20$, respectively; panel (d) shows the final hard
partition.}
\label{fig:antenna_examples2}
\end{figure*}
Figure~\ref{fig:antenna_examples2} illustrates the
candidate-generation process for a selected channel realization
in which the final search over $K\in\{25,\ldots,35\}$ chooses
$K=30$.
For this illustration, the $K=30$ branch starts from a
regularly dispersed partition.
Twenty updates are performed at each of the two penalty
levels, $\mu=0.01$ and $0.05$.
The intermediate panels display continuous optimization
weights.
The selected
partition contains 24 boundary compensation antennas,
compared with 14 in the initial configuration.
This example illustrates spatial redistribution toward the
array boundary when PA distortion dominates.

\begin{figure}[t]
  \centering
  \includegraphics[]{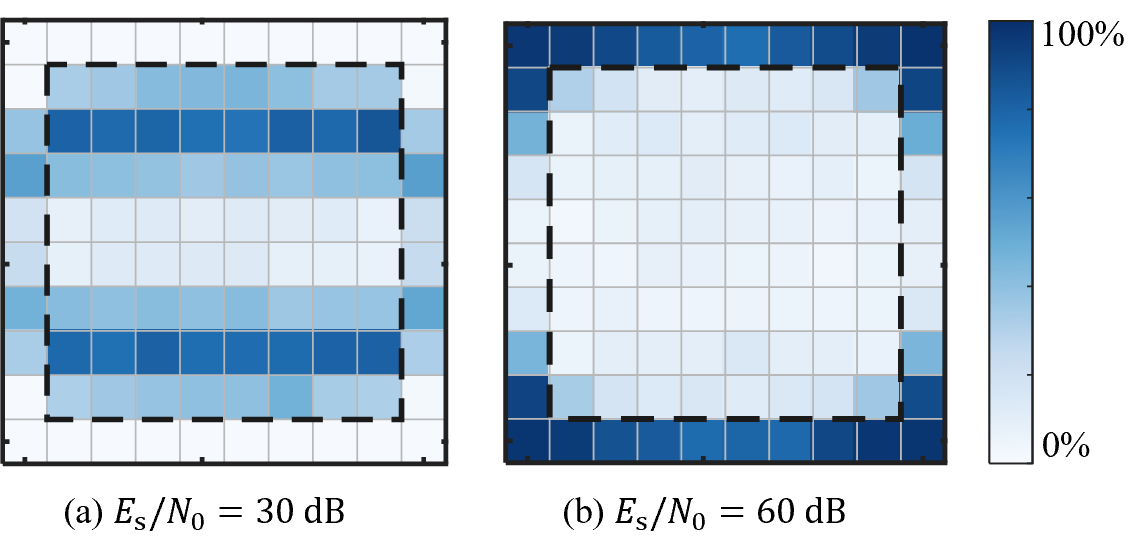}
  \caption{Empirical in-band distortion compensation-antenna selection probabilities for Proposed Full at $E_{\mathrm{s}}/N_0=30$ and 60~dB. The dashed square separates the outermost ring from the interior elements.}
  \label{fig:selection_heatmap}
\end{figure}
Figure~\ref{fig:selection_heatmap} shows that the preferred positions depend on $E_\mathrm s/N_0$. At 60~dB, where PA distortion dominates, boundary antennas are selected in most
trials, in line with Outer Placement~\cite{Li_ACCESS}. At 30~dB, the selections concentrate on the
interior, leaving the outermost elements mostly to data
transmission.

\begin{figure}[t]
  \centering
  \includegraphics[width=\linewidth]{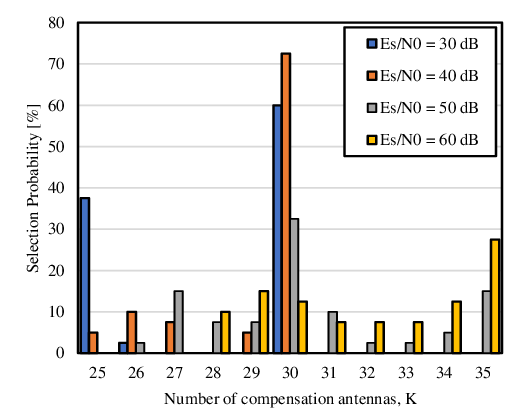}
  \caption{Empirical distributions of the compensation-antenna
  count selected by Proposed Full at different $E_{\mathrm{s}}/N_0$ values. Each
  distribution contains 40 trials, with candidate counts
  $K\in\{25,26,\ldots,35\}$.}
  \label{fig:k_distribution}
\end{figure}
The count distributions in Fig.~\ref{fig:k_distribution} show a
related adaptation. At 30~dB, $K=25$ and $K=30$ account for
37.5\% and 60\% of the trials, respectively. At 60~dB, the
selected count spans from 28 to 35, with $K=35$ occurring in 27.5\%
of the trials. The results are consistent with a channel-dependent
tradeoff between preserving data antenna array resources and
allocating degrees of freedom to compensation.

\begin{figure}[t]
  \centering
  \includegraphics[width=\linewidth]{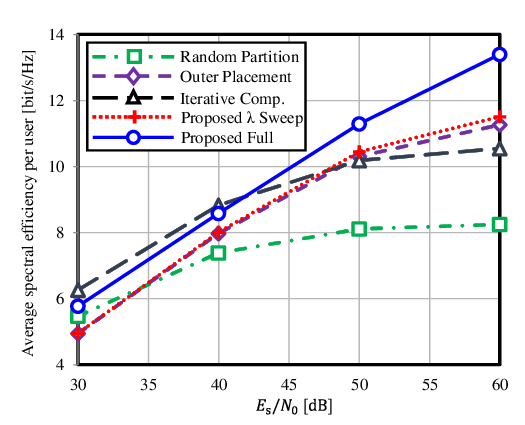}
\caption{Average SIDNR-based spectral efficiency per user
versus $E_{\mathrm{s}}/N_0$.}
  \label{fig:sumrate_comparison}
\end{figure}

\red{Figure~\ref{fig:sumrate_comparison} compares the average
SIDNR-based spectral efficiency per user computed using
\eqref{eq:measured_rate_per_user}.
Iterative compensation achieves the highest spectral
efficiency at low $E_{\mathrm{s}}/N_0$, whereas Proposed Full
performs best at 50 and 60~dB.
At 60~dB, Proposed Full achieves approximately
13.4~bit/s/Hz per user, compared with 11.5 for
Proposed $\lambda$ Sweep and 11.3 for Outer Placement.
Thus, the additional partition optimization improves
spectral efficiency by approximately 16\% over the inner-loop-only design at this operating point.}

\begin{table}[t]
\centering
\caption{Runtime and BER comparison. BER is evaluated over 200 waveform-level trials at $E_{\mathrm{s}}/N_0=60$~{\rm d}B.}
\label{tab:runtime_ber_comparison}
\footnotesize
\begin{tabular}{lcc}
\hline
Scheme & Runtime [s] & BER \\
\hline
Iterative Comp. (3 iterations)  & 38.61  & $2.6\times10^{-2}$ \\
Iterative Comp. (7 iterations)  & 81.09  & $1.2\times10^{-2}$ \\
Iterative Comp. (11 iterations) & 123.99 & $8.3\times10^{-3}$ \\
Iterative Comp. (20 iterations) & 219.22 & $5.8\times10^{-3}$ \\
Iterative Comp. (30 iterations) & 328.29 & $4.7\times10^{-3}$ \\
Outer Placement                 & 13.00  & $3.1\times10^{-3}$ \\
Proposed $\lambda$ Sweep        & 17.59  & $5.8\times10^{-4}$ \\
Proposed Full                   & 102.56 & $2.2\times10^{-5}$ \\
\hline
\end{tabular}
\end{table}

\subsection{Computational Cost}
Using full covariance matrices and treating each scalar
PC/PA moment evaluation as a constant cost,
one analytical objective evaluation costs
$C_{\mathrm a}=\mathcal O(|\mathcal L|(M^2U+MU^2+U^3))$,
where $|\mathcal L|$ is the number of occupied subcarriers.
The main computational cost comes from constructing the precoders and evaluating the received distortion
covariances.
Searching $N_\lambda=|\mathcal G_\lambda|$ regularization
candidates costs $\mathcal O(N_\lambda C_{\mathrm a})$,
while a coordinatewise finite-difference gradient at fixed
$\lambda_{\mathrm e}$ costs $\mathcal O(MC_{\mathrm a})$.

For each shortlisted partition, let $C_{\mathrm{PC}}$ denote the one-time data-waveform and PC preparation, including PC sample covariances, and let $C_{\mathrm{wf}}$ denote one regularization-dependent compensation-waveform, PA-statistics, and SIDNR evaluation. Conservative bounds are $C_{\mathrm{PC}}=\mathcal{O}\big(|\mathcal{L}|(MU^2+U^3+N_{\mathrm{sym}}M^2)+N_{\mathrm{sym}}MN\log N+N_{\mathrm{sym}}MN_{\mathrm{p}}^{\max}L_{\mathrm{pc}}\big)$ and $C_{\mathrm{wf}}=\mathcal{O}\big(C_{\mathrm{a}}+N_{\mathrm{sym}}|\mathcal{L}|M^2+N_{\mathrm{sym}}MN\log N\big)$. Here, $N_{\mathrm{sym}}$ is the number of OFDM symbols per trial, $N$ is the DFT size, $N_{\mathrm{p}}^{\max}$ bounds the number of PC pulses added per antenna and symbol, and $L_{\mathrm{pc}}$ is the PC pulse length in samples. In this evaluation, the full-symbol circular kernel gives $L_{\mathrm{pc}}=N$.

With bounded backtracking, at most $T_q$ continuous updates per candidate $K$ across all penalty stages, and a shortlist $\mathcal{K}_{\mathrm{s}}$ of $N_{\mathrm{s}}=|\mathcal{K}_{\mathrm{s}}|$ partitions, the total complexity is $C_{\mathrm{Full}}=\mathcal{O}\big(|\mathcal{K}|T_q(N_\lambda+M)C_{\mathrm{a}}+N_{\mathrm{s}}(C_{\mathrm{PC}}+N_\lambda C_{\mathrm{wf}})\big)$. It should be noted that waveform-dependent computation is restricted to the shortlisted partitions, whereas the computational cost of the analytical search is independent of $N_{\mathrm{sym}}$.

For Proposed $\lambda$ Sweep, the antenna partition is
fixed, and the complexity reduces to
$\mathcal O(C_{\mathrm{PC}}+N_\lambda C_{\mathrm{wf}})$,
since PC waveforms are prepared once and reused across
the regularization search.

Table~\ref{tab:runtime_ber_comparison} reports the runtime
and BER of the compared implementations.
The runtime of iterative compensation increases
approximately linearly with the number of iterations.
Proposed Full requires additional partition-search
time compared with Proposed $\lambda$ Sweep,
but achieves lower BER and shorter runtime than the 11-, 20-, and 30-iteration compensation schemes in the tested implementations.

\section{Conclusions}
\label{Sect:Conclusions}
This paper proposed a joint optimization method for antenna partitioning and distortion-aware compensation precoding in massive MIMO-OFDM systems with nonlinear PAs, where peak cancellation and in-band distortion compensation are employed. 
In the proposed method, a covariance-based analytical framework for joint optimization was developed by combining PC and PA distortion statistics with within-branch spatial covariance approximations to derive a per-user, per-subcarrier SIDNR approximation for transmitter design. 
Under a fixed PC threshold and ZF data precoding, the proposed layered procedure jointly adapts the antenna partition and scalar-RZF compensation regularization. The practical design metric neglects additional data-branch PA distortion in the considered operating regime, whereas waveform-level evaluation retains nonlinear PAs on both branches. Our results demonstrated additional BER and SIDNR-based spectral-efficiency gains from joint partition and precoder adaptation, particularly in the high $E_{\mathrm{s}}/N_0$ regime where PA nonlinearity dominates the system behavior. These gains come with additional partition-search cost, while fixed-partition regularization optimization provides a lower-cost design option.

\section*{Acknowledgment}
This research was supported in part by JSPS Grants-in-Aid for Scientific Research 24K07490, and in part by the German Federal Ministry of Research,
Technology and Space within the 6GEM+ Transfer Hub under grant
16KIS2409K.


\begin{thebibliography}{99}

\bibitem{RahmatallahMohan2013}
Y.~Rahmatallah and S.~Mohan,
``Peak-To-Average Power Ratio Reduction in OFDM Systems: A Survey and Taxonomy,''
\emph{IEEE Communications Surveys \& Tutorials},
vol.~15, no.~4, pp.~1567--1592, Fourth Quarter 2013.

  \bibitem{Ochiai2013}
  H.~Ochiai,
  ``An Analysis of Band-limited Communication Systems from Amplifier Efficiency and Distortion Perspective,''
  \emph{IEEE Transactions on Communications}, vol.~61, no.~4, pp.~1460--1472, Apr.~2013.

\bibitem{Zayani2019PAPR}
R.~Zayani, H.~Shaiek, and D.~Roviras,
``PAPR-Aware Massive MIMO-OFDM Downlink,''
\emph{IEEE Access},
vol.~7, pp.~25474--25484, Feb.~2019.

\bibitem{SongOchiai2016}
  J.~Song and H.~Ochiai,
  ``Performance Analysis for OFDM Signals With Peak Cancellation,''
  \emph{IEEE Transactions on Communications}, vol.~64, no.~1, pp.~261--270, Jan.~2016.

  \bibitem{tvtKageyama}T. Kageyama, O. Muta, and H. Gacanin, ``Performance Analysis of OFDM With Peak Cancellation Under EVM and ACLR Restrictions,'' \textit{IEEE Transactions on Vehicular Technology}, vol. 69, no. 6, pp. 6230-6241, June 2020.

  \bibitem{KageyamaAccess2020}
T.~Kageyama, O.~Muta, and H.~Gacanin,
``Enhanced Peak Cancellation With Simplified In-Band Distortion Compensation for Massive MIMO-OFDM,''
\emph{IEEE Access},
vol.~8, pp.~73420--73431, 2020.

\bibitem{Li_VTC}
Z. Li, M. Omoto, O. Muta, and K. Maruta, ``In-band Distortion-aware MMSE-Precoding for Massive MIMO-OFDM with Peak Cancellation,'' IEEE Vehicular Technology Conference Fall 2025, Oct. 2025.

\bibitem{Li_ACCESS}
Z. Li, R. Nojima, O. Muta, M. Omoto, T. Makita, and K. Maruta, ``In-Band Distortion Compensation for Peak Cancellation-Based PAPR Reduction in Massive MIMO-OFDM With Uniform Planar Array,'' \textit{IEEE Access}, vol. 12, pp. 156118-156130, 2024.

\bibitem{Li_IEICE}
Z.~Li, K.~Inada, M.~Omoto, O.~Muta, K.~Ikuta, and C.-H.~Huang,
``Performance Evaluation of Peak Cancellation with Antenna-Placement-Aware In-Band Distortion Compensation for Subarray-Based Hybrid Massive MIMO-OFDM Systems,''
\emph{IEICE Transactions on Communications}, accepted for publication, 2026.

\bibitem{Prabhu2014PAPR}
H.~Prabhu, O.~Edfors, J.~Rodrigues, L.~Liu, and F.~Rusek,
``A low-complex peak-to-average power reduction scheme for OFDM based massive MIMO systems,''
in \emph{Proc. IEEE International Symposium on Communications, Control and Signal Processing (ISCCSP)},
pp.~114--117, May 2014.

\bibitem{Khan2021TransmitDistortion}
M.~Q.~Khan, M.~D.~Nisar, and H.~A.~Khan,
``On PAPR Reduction and Transmit Distortion Compensation in Massive MIMO OFDM Systems,''
in \emph{Proc. IEEE Vehicular Technology Conference (VTC-Spring)},
pp.~1--5, Apr.~2021.

\bibitem{Saito2024ParallelPC}
J.~Saito, T.~Hara, and K.~Higuchi,
``Efficient Parallel PAPR Reduction Method Using Null-Constrained
Peak Cancellation Signal for Massive MIMO-OFDM Systems,''
\emph{IEEE Open Journal of the Communications Society},
vol.~5, pp.~3544--3555, 2024.

\bibitem{Muneer2022AntennaReservation}
S.~Muneer, J.~R.~Sanchez, L.~Van der Perre, O.~Edfors,
H.~Sj{\"o}land, and L.~Liu,
``System Design and Performance for Antenna Reservation in Massive MIMO,''
in \emph{Proc. IEEE 96th Vehicular Technology Conference
(VTC2022-Fall)}, London, U.K., Sept.~2022.

\bibitem{Muneer2026Clipping}
S.~Muneer, H.~Sj{\"o}land, O.~Edfors, and L.~Liu,
``Low Complexity Clipping Distortion Compensation for PAPR Reduction
in Massive MIMO-OFDM for Frequency Selective Channels,''
\emph{IEEE Transactions on Green Communications and Networking},
vol.~10, pp.~966--977, 2026.


\bibitem{SanayeiNosratinia2004}
S.~Sanayei and A.~Nosratinia,
``Antenna Selection in MIMO Systems,''
\emph{IEEE Communications Magazine},
vol.~42, no.~10, pp.~68--73, Oct.~2004.


  \bibitem{FDD} A. Nasser, M. Elsabrouty, and O. Muta, ``FDD Cooperative Channel Estimation and Feedback for 3D Massive MIMO System,'' \textit{IEEE Access}, vol. 7, pp. 76283-76294, June 2019.


\bibitem{Peel2005RZF}
C.~B.~Peel, B.~M.~Hochwald, and A.~L.~Swindlehurst,
``A vector-perturbation technique for near-capacity multiantenna multiuser communication---Part I: Channel inversion and regularization,''
\emph{IEEE Transactions on Communications},
vol.~53, no.~1, pp.~195--202, Jan.~2005.

\bibitem{Bussgang1952}
J.~J.~Bussgang,
``Crosscorrelation Functions of Amplitude-Distorted Gaussian Signals,''
Research Laboratory of Electronics, Massachusetts Institute of Technology,
Technical Report~216, Mar.~1952.



  \bibitem{BillingsleyPM}
P.~Billingsley, \emph{Probability and Measure}, 3rd~ed.
New York, NY, USA: Wiley, 1995.


\bibitem{Hoffmann2026PA}
M.~Hoffmann and P.~Kryszkiewicz,
``Impact of Nonlinear Power Amplifier on Massive MIMO:
Machine Learning Prediction Under Realistic Radio Channel,''
\emph{IEEE Transactions on Vehicular Technology},
vol. 75, no. 9, pp. 20072-20087, Sept. 2026.


\bibitem{Rapp1991}
C.~Rapp,
``Effects of HPA-nonlinearity on a 4-DPSK/OFDM-signal for a digital sound broadcasting system,''
in \emph{Proc. 2nd European Conference on Satellite Communications},
pp.~179--184, Oct.~1991.


\end{thebibliography}
\end{document}